\documentclass[a4paper,twocolumn]{article}
\usepackage{amssymb,amsmath}
\usepackage{graphics}
\usepackage{maybemath}
\usepackage[pdftex]{xcolor}
\usepackage{hyperref}
\usepackage[numbers,square,sort&compress]{natbib}

\title{Shell and explosive hydrogen burning}
\date{}

\author{
A. Boeltzig \\ 
Gran Sasso Science Institute, L'Aquila (Italy) \\ \and
C.G. Bruno \\  
SUPA, School of Physics and Astronomy, University of Edinburgh,  Edinburgh (UK) \\ \and %
F. Cavanna \\  Dipartimento di Fisica Universit\`a di Genova, and \\ INFN Sezione di Genova, Genova (Italy) \\  \and%
S. Cristallo \\ INAF, Osservatorio Astronomico di Collurania, Teramo (Italy) and \\ INFN Sezione di Napoli, Napoli (Italy) \\  \and %
T. Davinson \\  
SUPA, School of Physics and Astronomy, University of Edinburgh,  Edinburgh (UK) \\   \and %
R. Depalo \\ Dipartimento di Fisica e Astronomia Universit\`a di Padova, Padova (Italy) and \\ INFN Sezione di Padova, Padova (Italy) \\ \and %
R.J. deBoer \\ Institute for Structure and Nuclear Astrophysics, Joint Institute for \\ Nuclear Astrophysics, University of Notre Dame, Notre Dame (Indiana, USA) \\  \and %
A. Di Leva \\ Dipartimento di Fisica Universit\`a di Napoli Federico II, Napoli (Italy) and \\ INFN Sezione di Napoli, Napoli (Italy) \and %
F. Ferraro \\ Dipartimento di Fisica Universit\`a di Genova, and \\ INFN Sezione di Genova, Genova (Italy) \\  \and%
G. Imbriani\thanks{\emph{Corresponding author:} gianluca.imbriani@unina.it}\\ Dipartimento di Fisica Universit\`a di Napoli Federico II, Napoli (Italy) and \\ INFN Sezione di Napoli, Napoli (Italy) \and 
P. Marigo \\ Dipartimento di Fisica e Astronomia Universit\`a di Padova, Padova (Italy)   \and%
F. Terrasi \\ Dipartimento di Matematica e Fisica Seconda Universit\`a di Napoli, Caserta (Italy) and \\ INFN Sezione di Napoli, Napoli (Italy)   \and%
M. Wiescher \\ Institute for Structure and Nuclear Astrophysics, Joint Institute for \\ Nuclear Astrophysics, University of Notre Dame, Notre Dame (Indiana, USA)%
}

\newcommand{\fifNpg}{^{15}{\rm N}(p,\gamma)^{16}{\rm O}}
\newcommand{\fifNpa}{^{15}{\rm N}(p,\alpha)^{12}{\rm C}}
\newcommand{\sevOpg}{^{17}{\rm O}(p,\gamma)^{18}{\rm F}}
\newcommand{\sevOpa}{^{17}{\rm O}(p,\alpha)^{14}{\rm N}}
\newcommand{\eigOpa}{^{18}{\rm O}(p,\alpha)^{15}{\rm N}}
\newcommand{\Nepg}{^{22}{\rm Ne}(p,\gamma)^{23}{\rm Na}}
\newcommand{\Napg}{^{23}{\rm Na}(p,\gamma)^{24}{\rm Mg}}
\newcommand{\Mgpg}{^{25}{\rm Mg}(p,\gamma)^{26}{\rm Al}}
\begin{document}
\thispagestyle{empty}
\clearpage\maketitle

\begin{abstract}
The nucleosynthesis of light elements, from helium up to silicon, mainly occurs in Red Giant and Asymptotic Giant Branch stars and Novae. The relative abundances of the synthesized nuclides critically depend on the rates of the nuclear processes involved, often through non-trivial reaction chains, combined with complex mixing mechanisms.
In this review, we summarize the contributions made by LUNA experiments in furthering our understanding of nuclear reaction rates necessary for modeling nucleosynthesis in AGB stars and Novae explosions.
\end{abstract}

\section*{Introduction}
\label{sec:Intro}
\begin{figure*}
\begin{center}
\resizebox{.9\textwidth}{!}{%
  \includegraphics{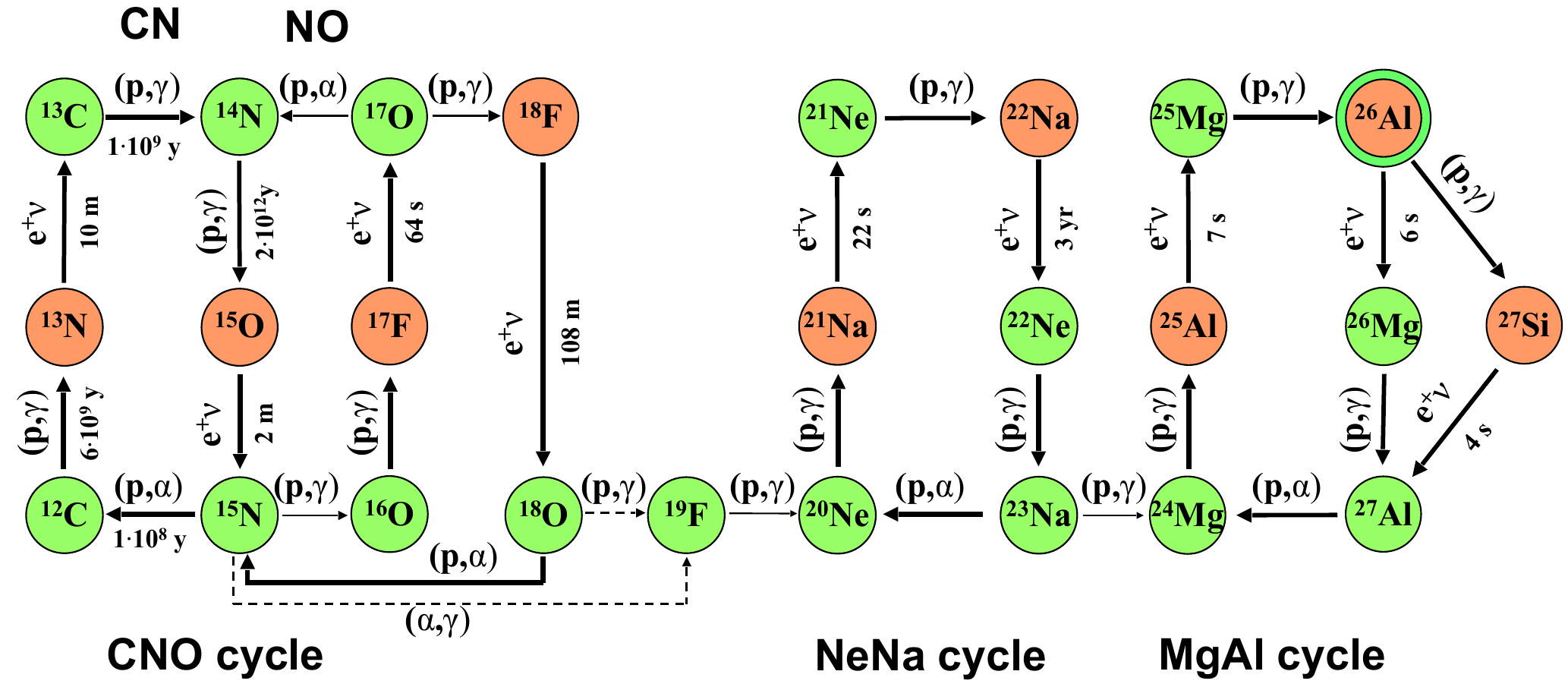}
}
\end{center}
\caption{Reaction network of CNO, NeNa, and MgAl cycles. Stable and long lived radioactive nuclei are shown in green, while short lived are shown in orange. $^{26}{\rm Al}$ is long lived nuclide in its ground state, while its metastable state decays directly to $^{26}{\rm Mg}$ with a short half life, see text for details. Characteristic time scales and lifetimes are indicated for some of the processes.}
\label{fig:cycles}       
\end{figure*}

In stars, after central hydrogen has been converted into helium, proton burning continues in a shell surrounding the core. Depending on the star's initial mass, this burning can proceed either through the $p$-$p$ chain or through the CNO cycle. The synthesis of heavier elements up to Si can occur through the NeNa and the MgAl cycles, as shown in Fig. \ref{fig:cycles}, provided that large enough temperatures are attained. A detailed knowledge of the nuclear processes active in these cycles is mandatory to properly determine the nucleosynthesis during the star evolution, in particular in the Red Giant Branch (RGB) and Asymptotic Giant Branch (AGB) phases, as well as in Novae explosions.
The abundances of the synthesized elements critically depend on the rates of the nuclear processes involved, often through non-trivial nucleosynthesis reaction chains, combined with complex mixing mechanisms.
In this review, we will summarize how the measurements performed at the Laboratory for Underground Nuclear Astrophysics (LUNA) contributed to the field.
As discussed earlier in this Topical Issue \cite{Best2016}, the Gran Sasso underground laboratory offers a unique low background environment where extremely weak processes can be studied to the precision needed for a fruitful comparison with astronomical observations.

The paper is organized as follows: in the first section the state of the art models of shell and explosive hydrogen burning are shortly outlined; Section \ref{sec:NuclearPhysics} recalls the basic tools of experimental nuclear astrophysics, especially in conjunction with the issue of extrapolating cross sections towards stellar energies; Section \ref{sec:Reactions} presents a review of reactions relevant for AGB and Novae nucleosynthesis studied at LUNA: the experimental results and the main implications for $\rm ^{15}N(p,\gamma)^{16}O$, $\rm ^{17}O(p,\gamma)^{18}F$, $\rm ^{17}O(p,\alpha)^{14}N$, $\rm ^{22}Ne(p,\gamma)^{23}Na$, $\rm ^{25}Mg(p,\gamma)^{26}Al$ are all discussed. Some upcoming projects, experiments in their advanced stages of planning and that are about to begin data collection, are also presented. 

\section{Astrophysics background}
\label{sec:AstroIntro}
Stars burn hydrogen via two different sets of nuclear reactions: the proton–-proton chain and the CNO cycle. 
The latter is the dominant nucleosynthetic path (and thus energy source) in stars with masses larger than about $\rm1.2\,M_\odot$.
Unlike the $p$-$p$ chain, the CNO cycle is a catalytic cycle, i.e. it converts 4 protons into one helium nucleus but does so via reactions on the preexistant seed nuclei of carbon, nitrogen and oxygen. Depending on the temperature attained in stellar interiors, different branches of the CNO cycle are activated, see Fig. \ref{fig:cycles}. For instance, at low temperatures (T$\sim 20\,$MK) only the CN cycle works, while at higher energies (T$\gtrsim 30\,$MK) the NO cycle is efficiently active as well.
The rate of the CN with respect to the NO cycle depends on the branching ratio of the proton capture on $^{15}$N, i.e. the $\fifNpg$ and $\fifNpa$ reaction cross sections. Indeed the probability for the first reaction to occur is about one for every thousand of the second one, thus the contribution to the overall nuclear energy production is negligible, while the consequences on the nucleosynthesis are critical.
Therefore, in case of an active NO cycle, the correct evaluation of the $\fifNpg$ reaction is crucial to properly predict the abundances of all the stable oxygen isotopes, ($^{16}$O, $^{17}$O and $^{18}$O) and their relative ratios. As a general rule, the abundance ratios of the oxygen isotopes are extremely sensitive to the temperature of the nucleosynthesis environment.
Therefore, the determination of proton capture rates on oxygen isotopes is of primary importance too, in particular for the $\sevOpg$ and the $\sevOpa$ reactions.\\*
The precise knowledge of the aforementioned nuclear processes is needed to address several astrophysical problems, spanning from RGB stars to Novae explosions, passing through the AGB phase.

The RGB phase starts at the so-called first dredge-up (FDU) \cite{Boothroyd1999}, a convective mixing episode occurring after the Main Sequence phase, that brings material from inner layers previously processed by CN cycling to the star's surface. 
As a consequence of the FDU, the surface C abundance decreases, as well as the $\rm^{12}C/^{13}C$ ratio, while the N abundance increases. 
The post-FDU oxygen isotopic ratios ($\rm^{16}O/^{17}O/^{18}O$) predicted by theoretical models lie on a characteristic line, their values depending on the initial stellar mass function, Ref. \cite{abia2012}.
However, the observed surface abundances of low mass giant stars ($\rm M<2\,M_\odot$) at the tip of the RGB on the Hertzsprung-Russell diagram often differ from those at FDU.  As an explanation for this difference, it has been proposed that the presence of a non-convective mixing episode links the surface to the hot layers above the H-burning shell. This occurs when stars populate the so-called bump of the luminosity function, Ref. \cite{palmerini2011} discusses the various proposed physical mechanisms triggering such a mixing.

The isotopic signatures found in pre-solar grains, meteoritic material of extra-solar origin, are a footprint of this nucleosynthesis. In particular, it has been suggested in Ref. \cite{Nittler1997} that Aluminum oxide grains (Al$_2$O$_3$) with a moderate $^{18}$O depletion and high $^{17}{\rm O}$ enrichment (Group 1), condensate in the outermost layers of RGB stars, when the C/O ratio is lower than 1. Models can reproduce Group 1 grain's isotopic ratios, only by assuming that the non-convective mixing mentioned above occurs during this evolutionary phase.
Even though various theories have been proposed to model such a mixing process, e.g. Ref. \cite{Nucci2014}, the differences between model predictions and  observations are still puzzling. Thus it is mandatory to reduce, as much as possible, the uncertainties affecting the nuclear processes of interest in order to better constrain the possible mixing phenomena.
\begin{figure*}[!htbp]
\begin{center}
\resizebox{.39\textwidth}{!}{%
  \includegraphics{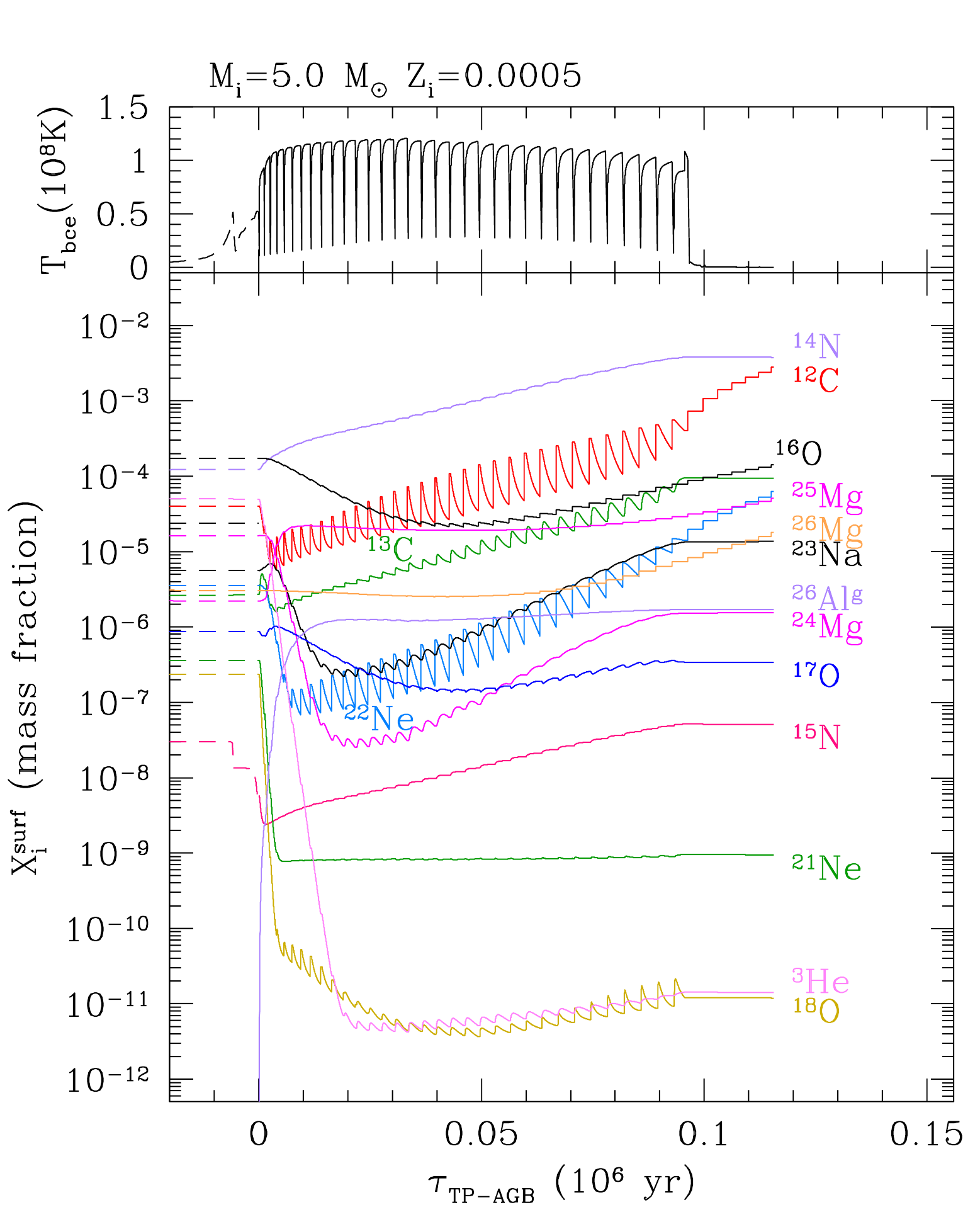}
}
\resizebox{.39\textwidth}{!}{%
  \includegraphics{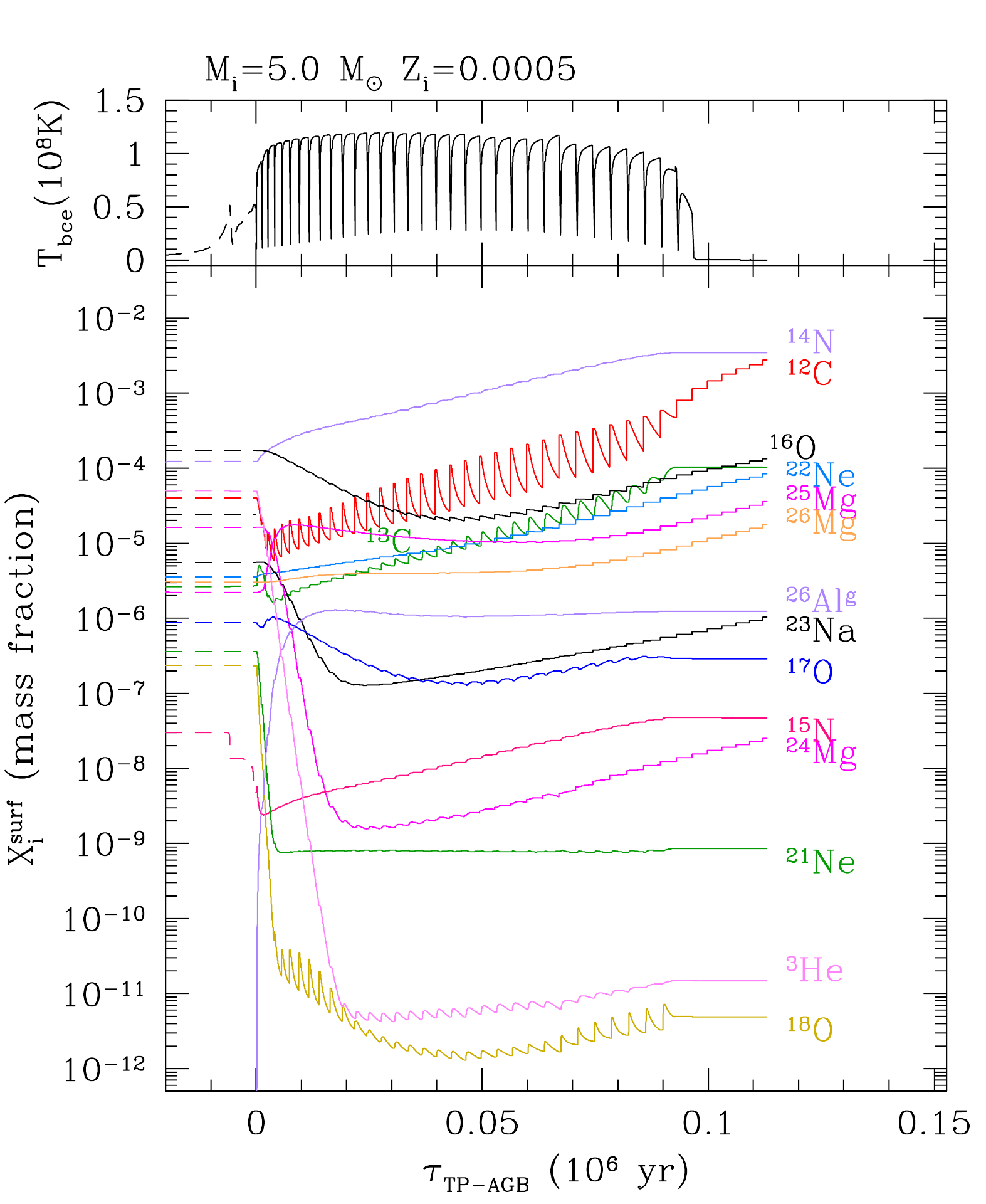}
}
\caption{Predicted evolution of the whole TP-AGB phase, up to the complete ejection of the envelope, of a star of $\rm5\,M_\odot$ initial mass and metallicity $Z_{\rm i} =0.0005$, computed with the COLIBRI code \cite{Marigo2013}. The TP-AGB star experiences recurrent TDU episodes as well as efficient HBB in which the CNO, NeNa, and MgAl cycles are at work.  The top panels show the temperature $T_{\rm bce}$ at the base of the convective envelope, The bottom panels present the surface abundances using different nuclear reaction rates. The calculation presented in the left panels uses rates as from \cite{NACRE99-NPA}, right panels with updated rates for $\fifNpg$ \cite{LeBlanc2010}, $\sevOpg$ \cite{DiLeva2014}, $\Nepg$ and $\Napg$ \cite{Iliadis2010}, $\Mgpg$ \cite{Straniero2013}.}
\label{fig:nucleosynthesis}
\end{center}
\end{figure*}

Al$_2$O$_3$ grains highly $^{18}$O depleted (Group 2) can be reproduced by models through the activation of a non-convective mixing in AGB stars as well \cite{palmerini2011}. 
During the AGB phase, the stellar structure consists of a partially degenerate carbon-oxygen core, a He shell, a H shell and a cool expanded convective envelope. The energy is mainly provided by the CNO proton burning in the H shell. 
This configuration is interrupted recurrently by the onset of a thermonuclear runaway (Thermal Pulse, TP) driven by the activation of the 3$\alpha$ process in the He-shell. When a TP occurs, the region between the two shells becomes unstable to convection for a short period, the external layers expand and the H shell burning temporarily stops.
As a consequence, the convective envelope penetrates into the C-rich and H-exhausted layers, bringing the freshly synthesized nucleosynthesis products to the surface. This is referred as Third Dredge Up (TDU), see e.g. Ref. \cite{Straniero2006}.
In sufficiently massive stars, $\rm M\gtrsim 5\,M_\odot$, depending on the metallicity, the inner region of the convective envelope is hot enough, $T\gtrsim60\rm\,MK$, to activate an efficient CNO cycling. This is known as Hot Bottom Burning (HBB), see e.g. Ref. \cite{Renzini1981}.
If this process is at work, the surface oxygen isotopic ratios could be modified. Unfortunately, due to the very high resolution needed, there are very few observations that can be used to significantly constrain the efficiency of HBB. However, as already said, the reduction of the uncertainties affecting the nuclear reactions that determine the equilibrium CNO abundances, is a necessary step in properly assessing the features of this process. As an example of the sensitivity of the nucleosynthesis on the nuclear reaction rates, Fig. \ref{fig:nucleosynthesis} shows the results of calculations for a $\rm 5\,M_\odot$ star performed with the COLIBRI evolutionary code \cite{Marigo2013}.

Another astrophysical site in which CNO plays a major role is the Novae explosion, see e.g. Refs. \cite{jose2007,Bode2008}. 
These phenomena usually take place in binary systems, consisting of a compact White Dwarf (WD) accreting H-rich material from a Main Sequence companion star. The WD has a degenerate CO or, in case of more massive objects, ONe core. When enough material has piled up on the WD surface, a thermonuclear runaway occurs leading to a convective mixing episode and, later, to the Nova phenomenon.
During that phase, temperatures as high as 400\,MK are attained at the base of the degenerate H-rich envelope, thus activating the so-called {\it hot} CNO cycle. Due to the very short convective turnover timescale, unstable isotopes (like $^{13}$N, $^{14,15}$O and $^{17,18}$F) can take part to the relevant nucleosynthesis at the WD surface.
In Ref. \cite{Iliadis2002} $\sevOpa$ was identified, among the aforementioned reactions, as the process whose current uncertainty most significantly affects the on-going nucleosynthesis.
The $\sevOpg$ reaction is important for the nucleosynthesis of $^{18}$F, as it powers a prompt $\gamma$-ray emission at 511\,keV from positron annihilation.
This radiation could in principle be detected by $\gamma$-ray telescopes and thus may represent a useful observable to check the nucleosynthesis in Novae explosions, although in practice its observation is extremely challenging \cite{Diehl2013}.

In massive AGB and super-AGB stars, initial masses $\rm\gtrsim 4 - 10\,M_{\odot}$, the temperature at the base of the convective envelope can be as high as $\rm T\simeq 60 - 100\,$MK, which leads to the activation of the NeNa and MgAl cycles \cite{Herwig_05, Karakas_etal06}, see Fig. \ref{fig:nucleosynthesis}.
During the last decade the nuclear reactions involved in these cycles, see Fig. \ref{fig:cycles}, received increasing attention. In fact they are believed to be the main agents of the observed anticorrelations in O-Na and Al-Mg  abundances exhibited by the stars of Galactic globular clusters \cite{Carretta_etal09, VenturaDantona_06}.
In particular, the production of $^{23}{\rm Na}$ is linked to the reaction $\Nepg$, whose cross section is still affected by large uncertainties. In Ref. \cite{Izzard2007} it was estimated that the current uncertainties in the nuclear reaction rates cause a typical variation of two orders of magnitude in the prediction of the global yields of $^{23}{\rm Na}$ from AGB stars experiencing HBB. In this respect, the predicted abundances of $^{22}{\rm Ne}$ and $^{23}{\rm Na}$ exhibit a remarkable difference, see Fig. \ref{fig:nucleosynthesis}.
The $\Nepg$ reaction plays a role also in Novae nucleosynthesis \cite{Jose1999}. Novae models predict that the $\Nepg$ reaction critically affects the final $^{22}$Ne abundance \cite{Iliadis2002},  and in turn the  $^{22}$Ne/$^{20}$Ne isotopic ratio \cite{Jose_etal04}.
Morever, CO novae models predict a significant effect on the $^{23}$Na and $^{24}$Mg yields in the ejecta \cite{Iliadis2002}.
$^{23}$Na can also be produced during carbon burning in massive stars and ejected in the Supernova explosion at the end of the star's life \cite{Limongi2006}.

During the pre-supernova stage, known as the Wolf-Rayet phase, massive stars having initial masses larger than about $10\,M_{\odot}$ experience large mass loss through intense stellar winds. The mass loss can be so efficient as to remove all the layers above those where $^{22}{\rm Ne}$ was converted to $^{23}{\rm Na}$ by H burning \cite{Abbott1987}.
To explain the observed trend in the $^{23}$Na abundances as a function of the metallicity, contributions from all nuclear processes and mixing mechanisms have to be taken into account \cite[][and references therein]{CunhaSmith_06,Gilli_etal06}.

Another key reaction is $\Mgpg$. In its ground state, the radioactive $\rm^{26}Al$ decays into the $E_{x}=1.8$\,MeV state of $^{26}$Mg, whose de-excitation produces a $\gamma$-ray of the same energy. This is at the origin of the diffuse Galactic emission at this energy, detected by the COMPTEL and INTEGRAL instruments \cite{Diehl1995,Bouchet_etal15}.
Several stellar sites are identified in the literature for the synthesis of $\rm^{26}Al$, namely: core collapse supernovae \cite{Limongi2006}, novae \cite{Jose1999}, massive AGB stars \cite{Mowlavi2000,Marigo2013},  and Wolf-Rayet stars \cite{Palacios2005}. 
Low mass AGB stars experiencing HBB may eject large quantities of $\rm^{26}Al$ through stellar winds \cite{Izzard2007,Straniero2013}. In this astrophysical site $^{26}$Al is expected to be produced efficiently in the quiescent H-burning shell during interpulse periods, and then brought up to the surface by the TDU.
This expectation is supported by the high $\rm^{26}Al/^{27}Al$ ratios measured in silicon carbide and corundum meteoritic grains \cite{Nittler1997}, as well as by the spectroscopic observation, in the nearby star IRC+10216, of transition lines that are likely to be produced by the $\rm^{26}AlCl$ and $\rm^{26}AlF$ molecules \cite{Guelin_etal95}.
The rate of the $\Mgpg$ reaction is expected to affect the isotopic ratios $\rm^{25}Mg/^{24}Mg$ and $\rm^{26}Mg/^{24}Mg$, that are accurately measured in presolar grains, and can be used to obtain information about the chemical evolution of the Galactic neighbourhood \cite{Kodolanyi_etal14}.

Interestingly, in Ref. \cite{Fenner_etal05} the effect of HBB nucleosynthesis on Mg isotopes has been studied in relation to what appears to be a variation of the fine structure constant deduced from quasar absorption lines at redshift $z<2$, which is sensitive to the concentrations of the Mg isotopes.
Recent analyses, e.g. Ref. \cite{Melendez2007}, have also suggested that the $^{25,26}$Mg isotopic ratios can be used to constrain the Galactic halo formation timescale.

\section{Experimental Nuclear Astrophysics basics}
\label{sec:NuclearPhysics}
For the reactions among low mass nuclei, that make up the networks for energy production and nucleosynthesis during BBN, stellar hydrogen and helium burning, models now desire cross sections with improved uncertainties. Often the goal is to reduce by a factor of two or more current uncertainty levels. Such improvements require unprecedentedly sensitive experimental measurements and more sophisticated analysis techniques.

The energies of interest in stellar burning scenarios are given approximately by the Gamow energy
\begin{equation} \label{eq:Gamow_energy}
E_\text{G} = \left[\left(\frac{\pi}{\hbar}\right)\left(\text{Z}_1\text{Z}_2e^2\right)^2\left(\frac{\mu}{2}\right)(kT)^2\right]^{1/3}\enspace,
\end{equation}
\noindent and width
\begin{equation} \label{eq:Gamow_width}
\Delta_\text{G} = \frac{4}{\sqrt{3}}\sqrt{E_\text{G}kT}\enspace,
\end{equation}
\noindent where Z$_1$ and Z$_2$ are the charges of the two interacting particles, $\mu$ is the reduced mass, and $T$ is the temperature of stellar burning (e.g. Ref.~\cite{BookIliadis}). Typically, the Gamow energy is few tens of keV for stellar hydrogen burning, see Table \ref{tab:sites}. Because of the repulsive Coulomb barrier between the two charged particles, the cross sections at these energies are very difficult or even impossible to measure directly.
\begin{table*}[!htb]
\scriptsize
\begin{center}
\begin{tabular}{|lccccccc|}
\hline
Site & Initial mass & $T$  & \multicolumn{5}{c|}{$E_G$} \\
~ & $[\rm M_\odot]$ & [MK] & \multicolumn{5}{c|}{[keV]} \\
\hline
~ &                & \phantom{$\bigg|$} & $^{15}{\rm N}+p$ & $^{17}{\rm O}+p$ & $^{22}{\rm Ne}+p$ & $^{23}{\rm Na}+p$ & $^{25}{\rm Mg}+p$ \\
RGB and Low mass AGB stars          & 1 - 3 & ~20 - 60~ & ~30 - 70~ & ~35 - 75~ & ~40 - 85~ & ~45 - 90~ & ~50 - 100 \\
Intermediate and massive AGB stars & 3 - 8 & ~20 - 120 & ~30 - 110 & ~35 - 120 & ~40 - 135 & ~45 - 145 & ~50 - 155 \\ 
Novae &                                                         & 200 - 500 & 150 - 280 & 160 - 300 & 190 - 350 & 200 - 380 & 210 - 400 \\
\hline
\end{tabular}
\end{center}
\caption{Representative energy ranges of the presented nuclear processes in the different stellar environments relevant for shell and explosive hydrogen burning. The range is given as the $E_G$ calculated at the minimum and maximum temperatures given in the $T$ column.}
\label{tab:sites}
\end{table*}%
The strategy then is threefold: make measurements that extend as low in energy as possible, use higher energy data to extrapolate down to low energies, and use other nuclear processes, typically nucleon transfer reactions, to populate the states very close or below threshold and probe their properties such as lifetime, width, and branching ratios. Transfer measurements have proved critical for reactions with strong subthreshold and direct contributions. The most confident determination of the cross section at stellar energies combines all of these independent strategies, attempting to achieve a consistent result.

At LUNA, measurements have been made to unprecedentedly low energies and cross sections by using high intensity proton and alpha beams and an extremely low background environment. Some of these measurements extend even down to the Gamow energy ranges of stellar burning. Even so, such measurements are quite challenging, pushing the limits of capability. For this reason, it is extremely useful to check the consistency of these low energy measurements with extrapolations from higher energy data and subthreshold and Direct Capture (DC) contributions that are characterized mainly by transfer and lifetime measurements. To perform such a comparison, a reaction theory or a model must be used to calculate the cross section over a wide energy range. The most common method is to use phenomenological $R$-matrix \cite{Descouvemont2010} to fit all the compound nucleus data simultaneously and include transfer reaction data by way of asymptotic normalization coefficients \cite{Mukhamedzhanov1999}.

\begin{figure}
\resizebox{0.45\textwidth}{!}{%
  \includegraphics{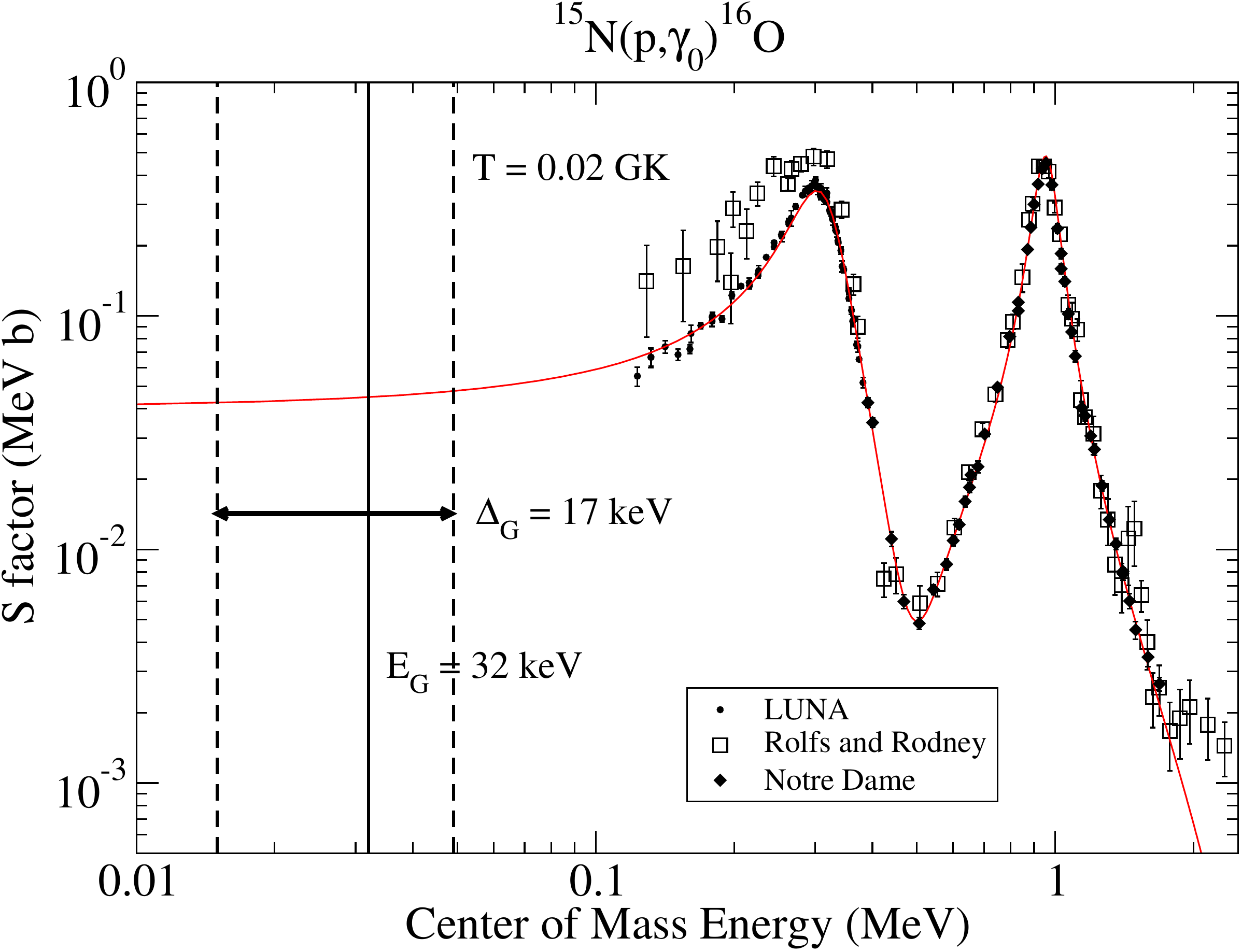}
}
\caption{Comparison of $S$-factor determinations for the reaction $^{15}$N$(p,\gamma_0)^{16}$O from Refs.~\cite{Rolfs1974450} and \cite{LeBlanc2010}. The Gamow energy and width are indicated for $T$ = 0.02\,GK, a typical temperature for H burning. The solid red line indicates the $R$-matrix fit.}
\label{fig:15N_pg_compare}       
\end{figure}

To illustrate, consider the reaction $^{15}$N$(p,\gamma)^{16}$O. Representative $S$ factors are shown in Fig.~\ref{fig:15N_pg_compare} from the measurement of Ref.~\cite{Rolfs1974450} and the more recent measurements made at LUNA and the University of Notre Dame Nuclear Science Laboratory (NSL) \cite{LeBlanc2010}. The $S$ factor is related to the cross section by
\begin{equation}
S(E) = E\sigma(E)e^{2\pi\eta}\enspace,
\end{equation}
\noindent where $\eta$ is the Sommerfeld parameter. Note the difference in the sensitivity of the two measurements as indicated by the error bars. The older measurement overestimates the cross section at low energy because of background contamination in the spectra, but this issue was overcome at LUNA by measuring in an underground environment.

\begin{figure}
\resizebox{0.45\textwidth}{!}{%
  \includegraphics{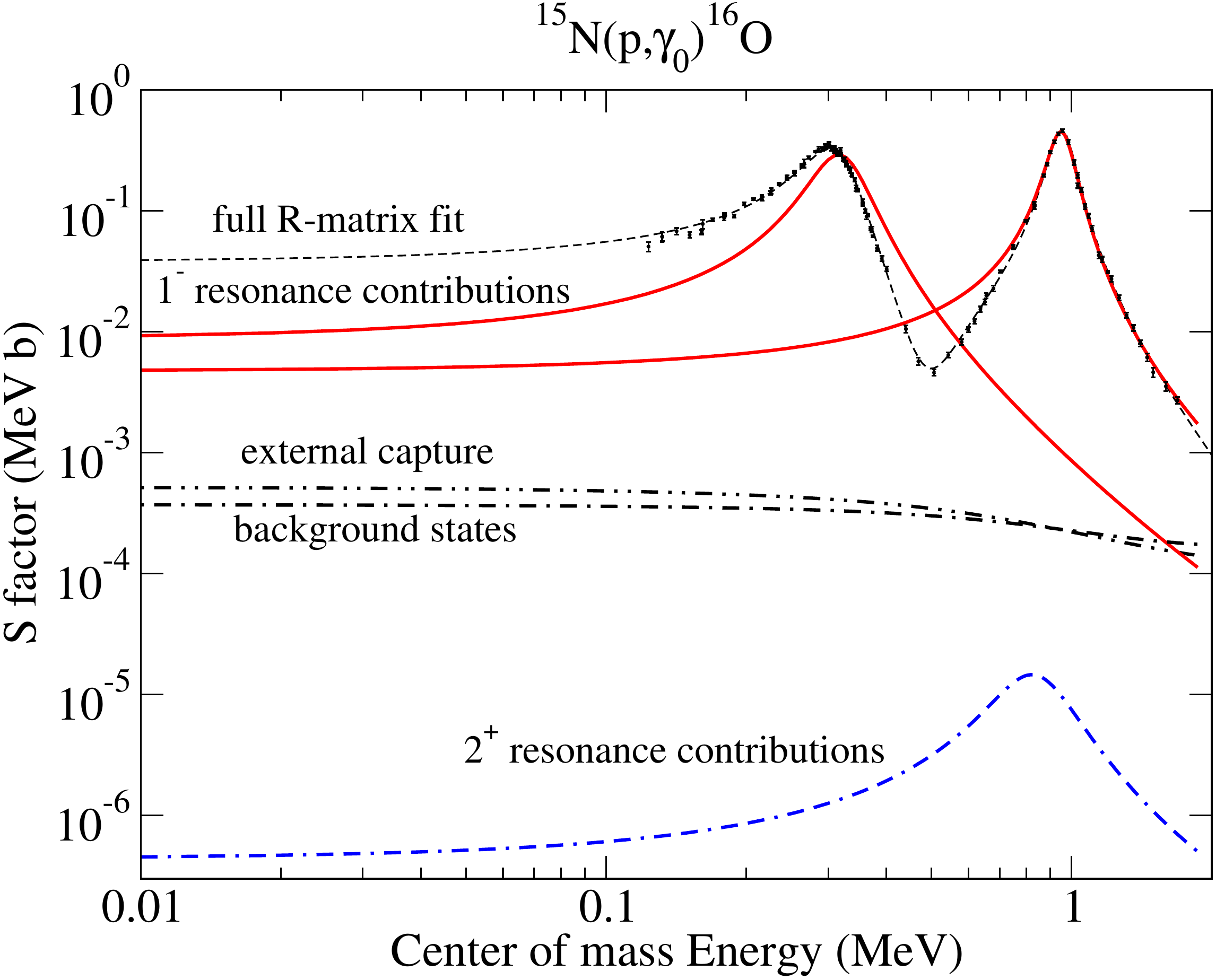}
}
\caption{The different $R$-matrix components used to model the $^{15}{\rm N}(p,\gamma_0)^{16}{\rm O}$ $S$-factor. The red solid lines represent the resonant contributions from the two broad $J^\pi$ = 1$^-$ levels, the blue dashed line from the $2^+$ level. While weak on its own, a DC component interferes with the resonant components strongly modifying the energy dependence in the off resonance regions.}
\label{fig:15Npg0_components_color}       
\end{figure}

Modeling the $^{15}$N$(p,\gamma)^{16}$O reaction is complicated by the presence of both resonant and DC contributions to the cross section. Fig.~\ref{fig:15Npg0_components_color} shows the break down of the $R$-matrix fit into these different contributions. Because of the level of complexity, having data over the entire energy range is critical. Further, transfer reaction studies have measured the proton asymptotic normalization coefficients ANCs of the ground state \cite{Mukhamedzhanov2008} placing strict limitations on the magnitude of the external capture contribution but the higher energy capture data must be used to determine the relative interference between the different contributions.

It must be emphasized how these three methods are mutually beneficial because there are strengths and weaknesses to all of them. Extrapolations from higher energy data have the advantage that the cross sections are larger making it much easier to gain high statistics and making angular distribution measurements tenable. However, with higher beam energies come more background reactions and covering a wider energy range requires more knowledge of the level structure of the compound nucleus to model. Measuring at very low energies is of utmost importance because it decreases the energy range of the extrapolation, on the other hand to achieve reasonable yields detectors must be placed in very close geometry thus necessitating assumptions about the reaction mechanisms in the analysis of the experimental results. The small count rates implies that background suppression is critical, which means that experiments can only be performed in underground facilities or by using complex active shielding techniques. Transfer reactions can provide ANCs that highly constrain the external capture contributions but their deduction from the experimental data have model dependencies whose uncertainties can be difficult to quantify. In the end, the ideal situation is to include as much data and structure information as possible in order to provide the higher consistency and reliability in the cross section determination. 

\section{Experimental achievements}
\label{sec:Reactions}

The LUNA facility at Gran Sasso, as described earlier in this Topical Issue \cite{Best2016}, operates a 400\,kV electrostatic accelerator, which delivers a proton beam of approximately 300\,$\mu$A and an $\alpha$ particle beam of about 200\,$\mu$A. 
Measurements generally last several weeks or even months making long term energy stability critical. Moreover good energy resolution is a fundamental requirement for narrow resonance studies and $S$-factor determinations. The LUNA 400\,kV accelerator has beam energy stability of 5\,eV/hour and the energy spread is of the order of 70\,eV \cite{Formicola2003}.

The LUNA collaboration uses a custom annular BGO detector that provides a large $\gamma$-ray detection efficiency at a moderate energy resolution. It is a valuable tool for detecting the sum signal of the $\gamma$-rays of the daughter nucleus' de-excitation, see e.g. \cite{Strieder2012}.
High sensitivity $\gamma$-ray spectroscopy is achieved using commercial High Purity Germanium (HPGe) detectors in close geometry, see e.g. \cite{Imbriani2005}. HPGe detectors provide superior $\gamma$-ray energy resolution, at the cost of a much lower detection efficiency.

In addition to background created by cosmic rays and environmental radioactive isotopes, beam induced background also has to be taken into account. Beam optics and target purity are among the most important ingredients for avoiding this additional source of background. 
In fact beam-induced background from contaminations of the target material can further hamper the detection of a reaction under study. For example, the broad resonance in $^{11}{\rm B}(p,\gamma)^{12}{\rm C}$ ($Q=16\rm\,MeV$) at $E_\mathrm{p} = 162\,\mathrm{keV}$ \cite{Ajzenberg1990}, can be an issue because of its high-energy $\gamma$-rays, with $^{11}{\rm B}$ content at a trace level. Hence a careful material selection and target preparation is mandatory in order to minimize contaminants content in the targets.

The LUNA collaboration has developed two beam lines, one is equipped with a gas target and the other with a solid target station. 
Gas targets have a higher purity and stability than solid targets at the price of technical complication \cite{Costantini2008}. Solid targets are generally made by evaporating, sputtering  or implanting the target nuclei into a backing material, see e.g. Ref. \cite{Caciolli2012}.
 
In the next subsections we report on the experiments that have been performed with the LUNA accelerator which are of interest for the shell and explosive hydrogen burning.

\subsection{The $\fifNpg$ reaction}
\label{sub:15Npg}
As shown in Fig. \ref{fig:cycles}, the $^{15}{\rm N}(p,\alpha)^{12}{\rm C}$ and $^{15}{\rm N}(p,\gamma)^{16}{\rm O}$ reactions form the bran\-ch point of the first CNO cycle. The CN cycle proceeds via:\\*
 $^{12}{\rm C}(p,\gamma)^{13}{\rm N}(e^+\nu)^{13}{\rm C}(p,\gamma)^{14}{\rm N}(p,\gamma)^{15}{\rm O}(e^+\nu)^{15}{\rm N}(p,\alpha)^{12}{\rm C}$. The $^{15}$N$(p,\alpha)^{12}$C always dominates over the  $^{15}$N$(p,\gamma)^{16}$O reaction rate, but a small fraction of the time the reaction network instead proceeds via:\\* $^{15}{\rm N}(p,\gamma)^{16}{\rm O}(p,\gamma)^{17}{\rm F}(e^+\nu)^{17}{\rm O}(p,\alpha)^{14}{\rm N}$\enspace.\\*
Because of this expanded network, the NO cycle, now produces the stable oxygen isotopes. The reaction rates ratio determines how much nucleosynthesis of $^{16,17,18}$O takes place during CNO burning \cite{Caughlan1962}.
\begin{figure}[!t]
\begin{center}
\resizebox{.9\hsize}{!}{\includegraphics{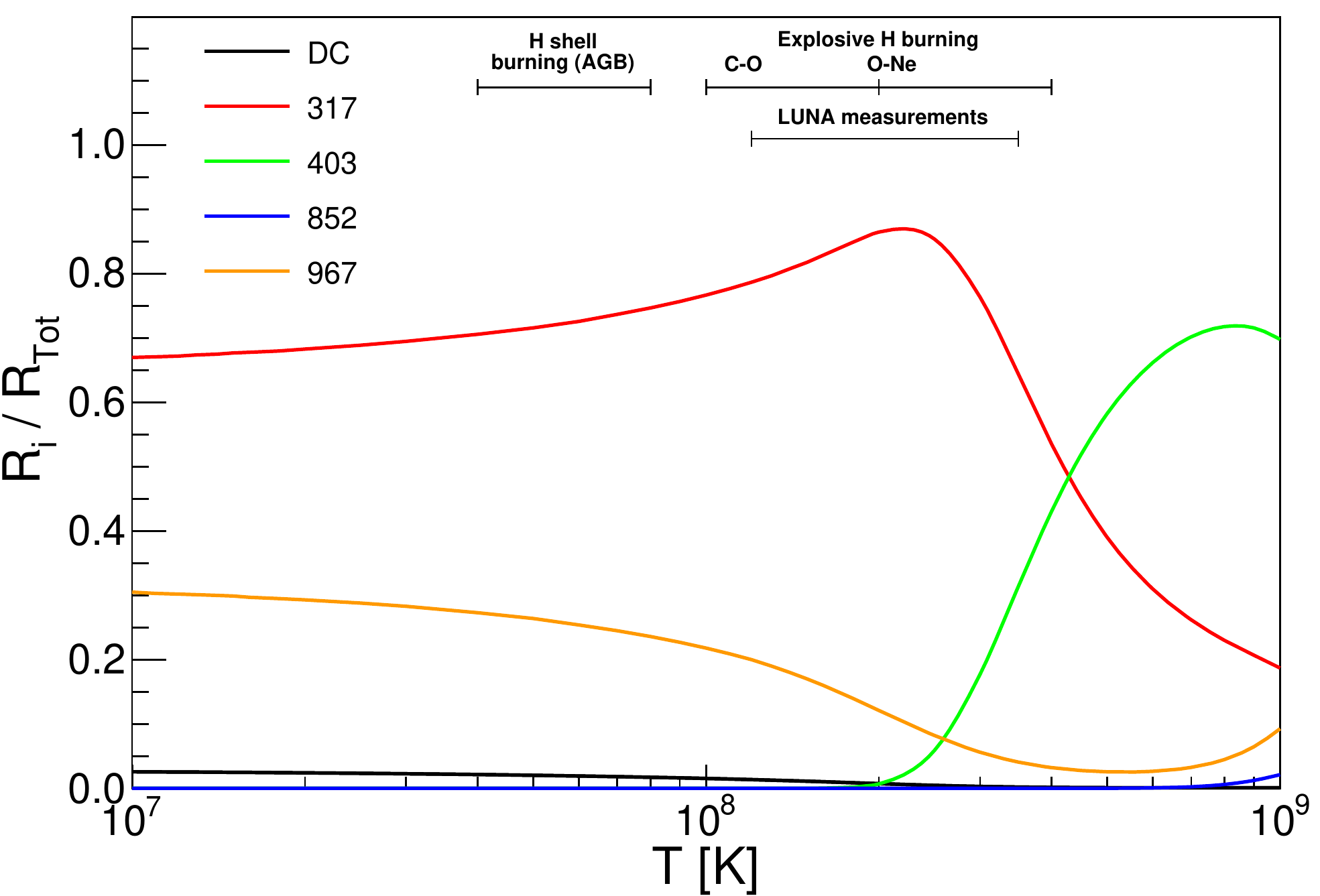}}
\caption{Fractional contribution of resonances and DC component to the total reaction rate of the $\fifNpg$, as a function of temperature. The resonances, except where otherwise indicated, are identified with their centre of mass energy in keV.
\label{fig:FractionalRate15Npg}}
\end{center}
\end{figure}

As described in Section \ref{sec:AstroIntro}, the astrophysically relevant temperatures are between 20 and 80\,MK. At these temperatures the reaction rate is dominated by the tails of the $E_{\rm c.m.}=317$ and 967\,keV broad resonances, see Fig. \ref{fig:FractionalRate15Npg}. In this figure the contribution of each resonance with respect to the total reaction rate is shown as a function of temperature. We find this figure useful to quickly understand which components are relevant at a given temperature, therefore a similar plot is present for all of the reactions that are discussed in this paper.

The $^{15}$N$(p,\gamma)^{16}$O reaction has been studied in great detail at LUNA. The earliest measurements are reported  in \cite{Bemmerer2009}, the total cross section was measured using the BGO detector setup over the energy range from $E_p$ = 90 to 230\,keV. This measurement extended to very low energies with low statistical uncertainty but suffered from large systematic uncertainties owing to $^{11}$B contamination in the target.  These data were expanded by a later measurement campaign \cite{Caciolli2011} with significantly reduced systematic uncertainties and a mapping over a somewhat wider energy range from $E_p$ = 72.8 to 368.3\,keV.

A joint measurement of the dominant ground state transition over a wide energy range from $E_p$ = 130 to 1800\,keV was then made at LUNA and at the University of Notre Dame \cite{LeBlanc2010}. This measurement confirmed the energy dependence of the original measurement by Ref. \cite{Hebbard1960289}, showing that Ref. \cite{Rolfs1974450} over predicted the cross section at low energies as shown in Fig. \ref{fig:15N_pg_compare}. Further, the measurements agreed with the $R$-matrix fits of Ref. \cite{Mukhamedzhanov2008}, motivated by proton ANC measurements, and Ref. \cite{Barker2008} based on the preference of $R$-matrix over interfering Breit-Wigner plus direct capture fits.

From a re-analysis of the $\gamma$ ray spectra of Ref. \cite{LeBlanc2010}, cross sections for the cascade transitions to the $E_x$ = 6.05, 6.13, and 7.12\,MeV bound states of $^{16}$O and the $^{15}$N$(p,\alpha\gamma_1)^{12}$C reaction were extracted \cite{imbriani2012}. An $R$-matrix analysis of the cascade transitions together with the ground state transition confirmed that the ground state dominates the cross section at stellar energies. The $^{15}$N$(p,\alpha\gamma_1)^{12}$C data were compared with that of Ref. \cite{Rolfs1974450} showing that the latter data were strongly distorted by nitrogen diffusion into the backing material. All of these results have been subsequently analyzed simultaneously using a global $R$-matrix fit in Ref. \cite{deBoer2013}, which demonstrated a high level of consistency between all the LUNA data and the proton ANCs of Ref. \cite{Mukhamedzhanov2008}.

The final fit of the $^{15}$N$(p,\gamma_0)^{16}$O data is shown in Fig. \ref{fig:15Npg0_components_color}. The cross section is dominated by two strong, interfering, 1$^-$ resonances. At low energy, which is in the off-resonance region, the cross section is also affected by the direct capture which is enhanced through interference with the resonant contributions. The strength of the direct capture is characterized in terms of the proton ANC and is fixed at the value given in Ref. \cite{Mukhamedzhanov2008}. A good fit also requires a contribution from higher lying states in the form of a background state. A 2$^+$ contribution is also present based on $^{12}$C$(\alpha,\gamma_0)^{16}$O data, but is below the sensitivity of current measurements. The ground state $S$-factor at zero energy was found to be 40(3)\,keV\,b \cite{deBoer2013}, in good agreement with Refs. \cite{Mukhamedzhanov2011}, 33.1-40.1\,keV\,b, and \cite{Barker2008},  $\approx$35\,keV\,b. The total $S$ factor is estimated at 41(3)\,keV\,b \cite{deBoer2013} owing to the small contributions from the cascade transitions \cite{imbriani2012}.

While the $^{15}$N$(p,\gamma)^{16}$O reaction can be considered now quite well known, it was shown in Ref. \cite{deBoer2013} that the data on this process can be used to constrain the high energy contributions to the cross section of the $^{12}$C$(\alpha,\gamma)^{16}$O reaction. In particular, further investigation of the cascade transitions over a wide energy range should prove very helpful. These transitions could be studied using either HPGe detectors, where the individual transitions can be separated, or using the BGO summing technique because at higher energies the cascade transitions begin to dominate the total cross section (see Fig.~4 of Ref. \cite{imbriani2012}). These experiments are ideal for the upcoming LUNA MV program \cite{Aliotta2016}.

\subsection{The $\sf\sevOpg$ reaction}
\label{sub:17Opg}
The $\sevOpg$ reaction, as described in Section \ref{sec:AstroIntro}, plays a key role in AGB nucleosynthesis, but it also has a large importance in explosive H burning occurring in type Ia novae. The energy range relevant for these two sites is at $E_{\rm c.m.}\lesssim500\,$keV.

\begin{figure}[!t]
\begin{center}
\resizebox{.9\hsize}{!}{\includegraphics{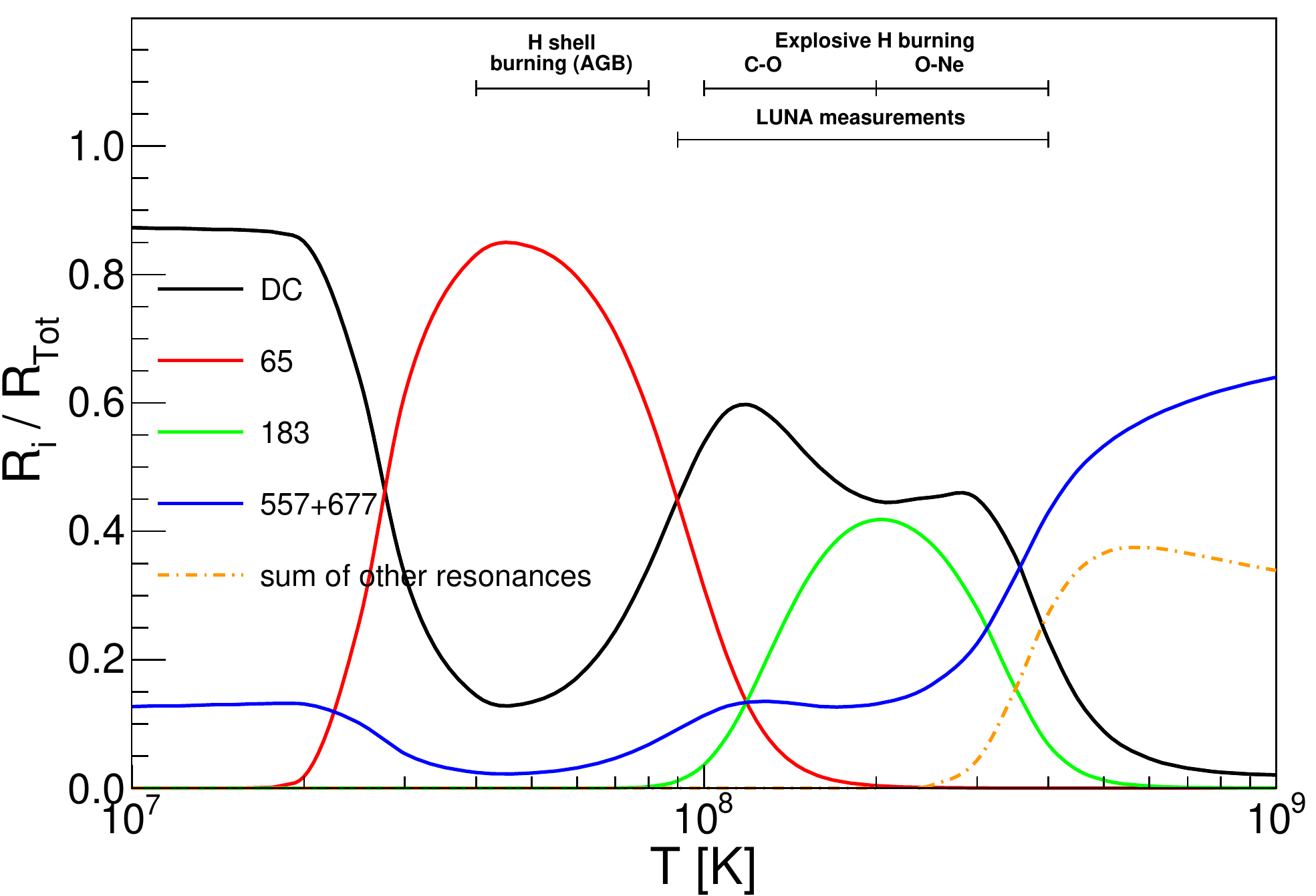}}
\caption{Fractional contributions to the reaction rate of $\sevOpg$, as a function of temperature.
\label{fig:FractionalRate17Opg}
}
\end{center}
\end{figure}

In the low energy range, the cross section is characterised  by two narrow resonances at $E_{\rm c.m.}=66$ and $183\,$keV and non-resonant contributions from the DC component and the tails of two broad resonances at $E_{\rm c.m.} = 557\,(J^\pi=3^+)$ and $677\,\rm keV (2^+)$. Figure \ref{fig:FractionalRate17Opg} shows the contribution to the reaction rate of the aforementioned components as a function of the temperature.

\noindent The temperature range $20\,{\rm MK}\lesssim T\lesssim80\,{\rm MK}$ is the most relevant to quiescent H burning and thus to AGB nucleosynthesis. In this interval the rate is mainly determined by the $E_{\rm c.m.}=66\,$keV resonance strength and to a lesser extent by the non-resonant component, see Fig. \ref{fig:FractionalRate17Opg}. The $E_{\rm c.m.}=66\,$keV resonance strength, $\omega\gamma=(19.0\pm3.2)\rm\,neV$, is presently determined only through indirect measurements \cite{Mak1980,Landre1989}. In a direct measurement this strength corresponds to about one reaction per delivered Coulomb on target, a rate still below the present measurement possibilities, but perhaps observable in the near future thanks to further background suppression, see Section \ref{sub:future}. 
The rate at temperatures relevant for novae nucleosynthesis, $100\,{\rm MK}\lesssim T\lesssim400\,{\rm MK}$, is determined by comparable contributions from the $E_{\rm c.m.}=183\,$keV resonance and the non-resonant components, see Fig. \ref{fig:FractionalRate17Opg}. 

The low energy range of the $\sevOpg$ process was investigated by several works either through prompt $\gamma$-ray detection, Refs. \cite{Rolfs1973,Fox2004,Fox2005,Newton2010,Kontos2012}, and more recently Ref. \cite{Buckner2015}, or through the off-line counting of $^{18}$F decays from irradiated targets \cite{Chafa2005,Chafa2007} (activation method). It was also measured through $^{18}$F recoils counting \cite{Hager2012}. The need of a further measurement of the cross section of this process was due to discrepancies present both in the $E_{\rm c.m.}=183\,$keV strength  and the non resonant contribution determinations. The measurements performed at LUNA have been extensively reported in \cite{Scott2012,DiLeva2014} while target production and characterisation are reported in \cite{Caciolli2012}. The measurements were performed using both the prompt $\gamma$-ray detection and the activation method. The measurement spanned the energy interval $E_{\rm c.m.}=160-370\,$keV. A proton beam of about 200$\,\mu$A intensity was delivered on solid $\rm Ta_2O_5$ targets. The prompt $\gamma$-ray detection was performed using a 120\% HPGe detector in close geometry, while activation measurements took advantage of the low background counting facilities present at LNGS \cite{Laubenstein2004}. 

The strength of the $E_{\rm c.m.}=183\,$keV resonance was determined to the highest precision to-date: $\omega\gamma=1.67\pm0.12\,\mu\rm eV$. At the resonance energy several previously unobserved transitions were also detected \cite{Scott2012}. The non-resonant $S$-factor in the energy region $100 - 450\rm\,keV$, most relevant for novae nucleosynthesis, was determined with high accuracy with a robust global fit taking into account all available literature data. The data were analyzed in a common fit procedure using a phenomenological approach \cite{Scott2012}. However, the upward slope in the S-factor at higher energies is the result of the tail of a broad resonance and the low energy dependence is dominated by direct capture. Therefore higher energy data is extremely helpful, if not necessary, in correctly fitting the data. Indeed the data analysis took advantage of the almost concurrent measurements of Ref. \cite{Kontos2012} that provided a precise determination of the parameters of the resonances at 557 and 677\,keV.
The astrophysical reaction rate, and the related uncertainty, was calculated in the appropriate temperature range using the approach of \cite{Iliadis2010_NPA1}. Reaction rates higher than previously recommended were found in the temperature region between 100 and $400\,$MK, mainly because of the higher $\omega\gamma$ of the $183\,$keV resonance \cite{Scott2012} and the larger non resonant contribution \cite{DiLeva2014}. Also the uncertainty on the reaction rate in the same temperature window was reduced by a factor of four. 
Certainly, a comprehensive $R$-matrix fit to high-energy and low-energy data could provide a more accurate extrapolation towards the AGB energy range.

The impact of the determination of the reaction rate on the expected nucleosynthesis products was performed by computing a set of nova models with the 1-D hydrodynamic SHIVA code \cite{Jose1998}: CO WDs with masses 1.0 and $1.15\rm\,M_\odot$ as well as ONe WDs with masses 1.15, 1.25 and $1.35\rm\,M_\odot$ were considered. The major outcome of this investigation was that the abundances of key isotopes such as $^{18}$F, $^{18}$O, $^{19}$F and $^{15}$N, although not very different (2-5\%) with respect to the use of previous reaction rate compilations, could be obtained with a precision of 10\%, i.e. sufficient to put firm constraints on observational features of novae nucleosynthesis \cite{DiLeva2014}.

\subsection{The $\sevOpa$ reaction}
\label{sub:17Opa}
\begin{figure}[!hbt]
\begin{center}
\resizebox{.9\hsize}{!}{\includegraphics{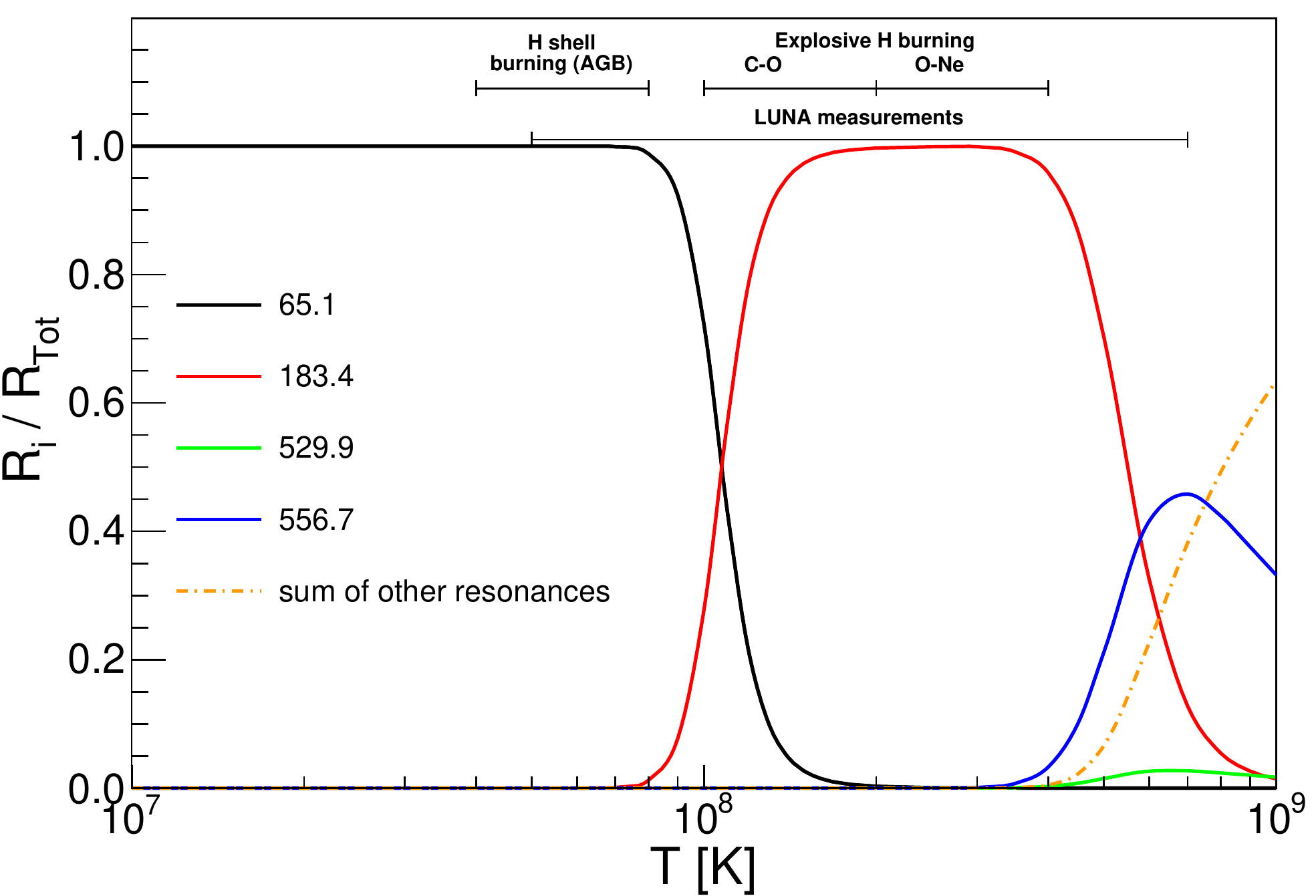}}
\caption{Fractional contribution of resonances to the total reaction rate of $\sevOpa$, as a function of temperature. 
\label{fig:FractionalRate17Opa}}
\end{center}
\end{figure}

At temperatures of astrophysical interest the reaction rate of the $\sevOpa$ reaction ($Q=1.2\,\rm MeV$) is dominated by a narrow and isolated resonance at $E_{\rm c.m.}$=65\,keV, as shown in Fig. \ref{fig:FractionalRate17Opa}. 
Previous measurements of this weak resonance, $\omega\gamma\sim$neV, have been attempted employing both direct \cite{blackmon1995} and indirect \cite{sergi2010} methods. In spite of this, the strength of this resonance remains uncertain by about 20\%.\\*
An experimental campaign aimed at precisely measuring the $\omega\gamma$ value of the $E_{\rm c.m.}$=65\,keV resonance in $\sevOpa$ was recently completed at LUNA. The thick-target yield technique was employed, i.e. protons were delivered on solid Ta$_2$O$_5$ targets, enriched in $^{17}$O. The $\alpha$ particles generated by the reaction ($E_\alpha\simeq1\,$MeV) were detected by eight silicon detectors placed at backward angles with respect to the beam axis.

The experimental setup was commissioned using the well-known resonance at $E_{\rm c.m.}=143\,$keV in $\eigOpa$\ and a resonance at $E_{\rm c.m.}=183.4\,$keV in $\sevOpa$. The latter resonance was especially useful for the commissioning, because the energy of the alpha particles emitted through it was extremely close to the energy expected for the alpha particles coming from the $E_{\rm c.m.}$=65\,keV resonance, allowing us to define precisely and reliably a region of interest for the signal. Only a few counts per hour were expected in typical experimental conditions for the 65\,keV resonance and having a region of interest defined {\it a priori} proved invaluable during the analyses.

The main contribution to the background in the region of interest of the signal is due to natural background. In order to quantify the reduction resulting from moving the silicon detectors underground a systematic, independent study was carried out. The background was measured underground and above ground with and without a lead shield mounted around the scattering chamber. A reduction of as much as a factor of 15 was observed in the region of interest between above ground and underground with this lead shield, confirming the advantage of moving underground for this measurement. Further details on the commissioning of the setup and the underground reduction in background for Silicon detectors are presented in Ref. \cite{Bruno2015}.\\
Thanks to the reduced background and the intense proton beam, a signal significantly above background could be observed during the measurement campaign for the 70\,keV resonance, data analysis is currently underway to estimate the resonance strength.

\subsection{The $\Nepg$ reaction}
\label{sub:22Nepg}
The $\Nepg$ reaction occurs in the NeNa cycle of CNO hydrogen burning, see Fig. \ref{fig:cycles}. This cycle plays a crucial role in the synthesis of the intermediate mass elements, as explained in greater detail in Section \ref{sec:AstroIntro}.

Among the reactions involved in the NeNa cycle, the one with the most uncertain cross section is $\Nepg$. In fact, according to the level structure of $^{23}$Na, several low-energy resonances are expected to contribute to the reaction rate at energies below 400\,keV. However, a large number of them, lying in the Gamow window \cite{Rolfs1988,Iliadis2010_NPA2}, have not been directly observed yet. These resonances have been investigated in previous experiments, but only upper limits on their strengths are reported \cite{Goerres82-NPA,Goerres83-NPA,Hale2001}. Resonances having better known strengths are the ones at $E_{\rm c.m.}$=458 and 1222\,keV \cite{Longland10-PRC,Keinonen1977}. 
Moreover, three resonances at $E_{\rm c.m.}$ = 68, 100 and 206\,keV have been tentatively reported in one indirect measurement \cite{Powers1971} but they have not been observed in more recent experiments \cite{Hale2001,Jenkins:2013fna}.
\begin{figure}[!t]
\begin{center}
\resizebox{.9\hsize}{!}{\includegraphics{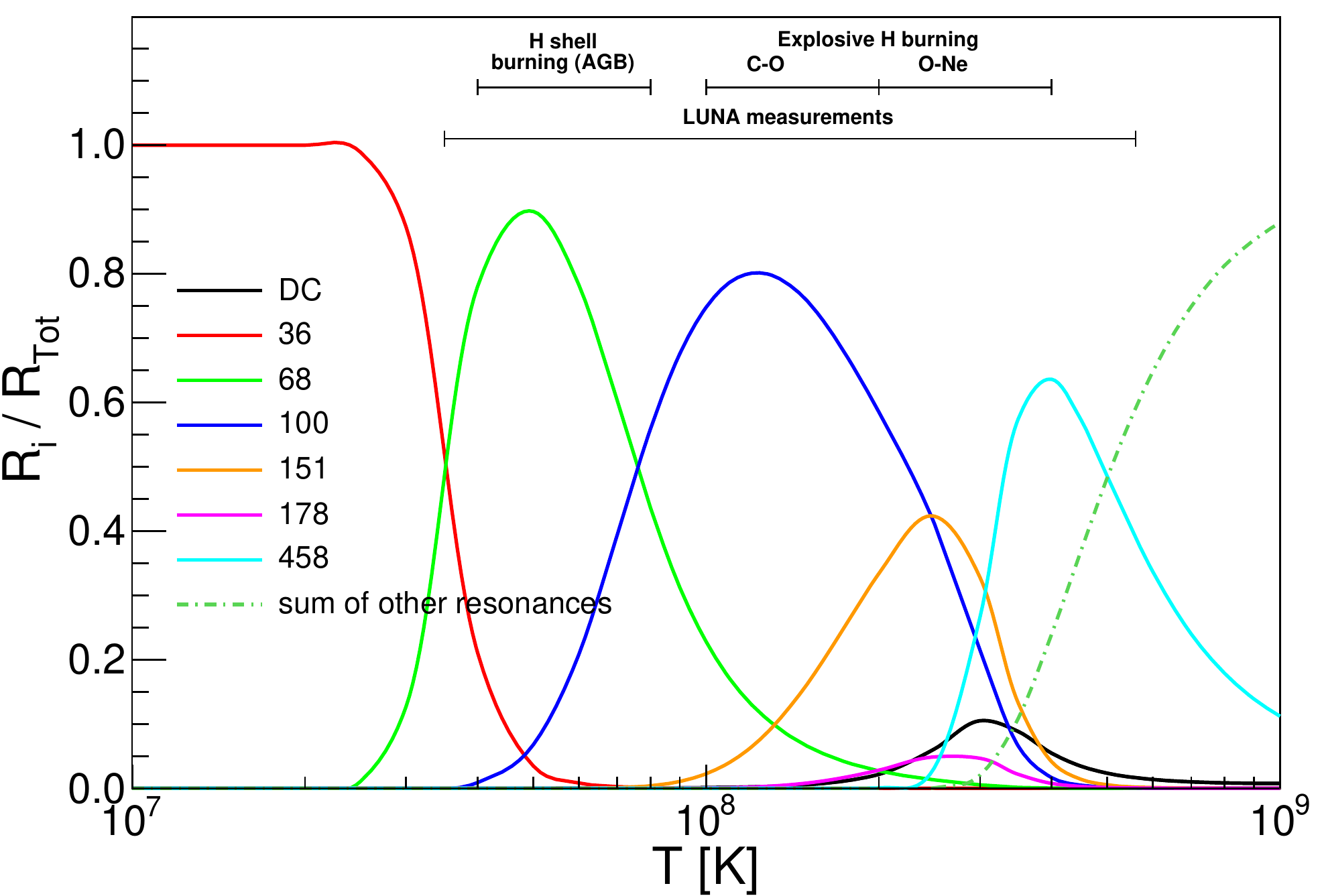}}
\caption{Fractional contributions to the reaction rate of $\Nepg$, as a function of temperature, according to prescriptions of Ref. \cite{NACRE99-NPA}.
\label{fig:FractionalRate22Nepg}}
\end{center}
\end{figure}
The complexity of this reaction is shown in Fig. \ref{fig:FractionalRate22Nepg}, where the contributions of the resonances believed to be significant to the total reaction rate are calculated according to Ref. \cite{NACRE99-NPA}.

\begin{figure}[!t]
\begin{center}
\resizebox{.9\hsize}{!}{\includegraphics{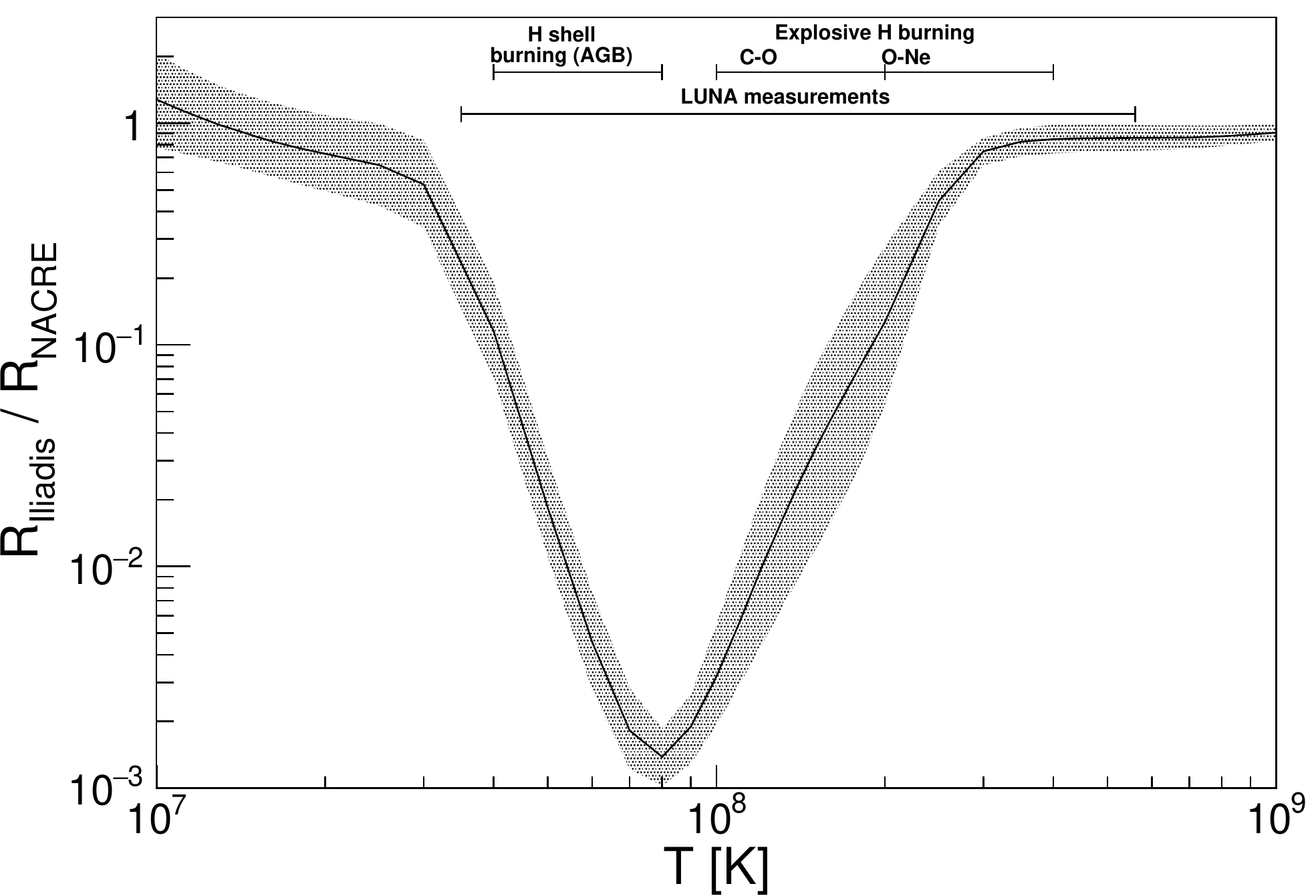}}
\caption{Ratio between the thermonuclear reaction rate calculated by Ref. \cite{Iliadis2010_NPA2} and NACRE \cite{NACRE99-NPA}.
\label{fig:RatioRate22Nepg}}
\end{center}
\end{figure}

So far the indirect data and the upper limits were treated in different ways in the literature, leading to large differences, e.g. a discrepancy of a factor of 1000 is present between the reaction rates reported in the NACRE compilation \cite{NACRE99-NPA} and the more recent evaluation of Ref. \cite{Iliadis2010_NPA2}, as shown in Fig. \ref{fig:RatioRate22Nepg}.
\noindent Sensitivity studies on AGB and classical novae nucleosynthesis, see e.g. Refs. \cite{Iliadis2002,Izzard2007}, show that  the uncertainty on the $\Nepg$ reaction rate significantly affects the predicted elemental abundances. Indeed changes on the yield of the isotopes between $^{20}$Ne and $^{27}$Al as large as two orders of magnitude are obtained depending on the choice on the reaction rate. 

A measurement campaign of the $\Nepg$ resonances has been recently completed at LUNA \cite{Cavanna2015}. The setup consisted of a windowless gas target with three differential pumping stages equipped with a gas recirculation and purification system \cite{Casella2002,Cavanna:2014lia}. The $\gamma$-ray detection was performed by means of two HPGe detectors, one at 90$^{\circ}$ and the other at 55$^{\circ}$ effective angle with respect to the beam direction. The two detectors were surrounded by massive copper and lead shielding in order to further suppress the environmental background. The use of high resolution detectors allowed both the measurement of the total resonance strength and the characterisation of the decay scheme of the resonances, which was still unknown for most of those investigated. Furthermore, the comparison of the counting rates of the two detectors allows for upper limits on effects due to $\gamma$-ray angular distributions to be determined.

The setup commissioning measurements, including gas density profile, beam heating effects, and beam induced background characterisation, are reported in \cite{Cavanna:2014lia}. These measurements were performed with neon gas of natural isotopic composition ($90.48\%$ $^{20}$Ne, $0.27\%$ $^{21}$Ne, $9.25\%$ $^{22}$Ne). Despite the low $^{22}$Ne content it was possible to directly observe for the first time the resonances at $E_{\rm c.m.}$=151, 178 and 248\,keV \cite{Cavanna2015}. For the resonance at $E_{\rm c.m.}$=178\,keV corresponding to the $E_x=8972\,$keV excited state in $^{23}$Na it was possible to set a lower limit on its strength of $\omega\gamma\geq 0.12\times10^{-6}\,\rm eV$ \cite{Cavanna:2014lia}, compared to the previous experimental upper limit of $2.6\times10^{-6}$\,eV \cite{Goerres82-NPA}. Because of its energy, this resonance has the largest impact on the reaction rate near 180\,MK, see Fig. \ref{fig:FractionalRate22Nepg}. At this temperature, with the current uncertainties, the resonance actual contribution to the total rate can change drastically. In fact, with respect to the rates reported in \cite{Iliadis2010_NPA2}, the contribution may change from 6\% of the total thermonuclear reaction rate, if the lower limit is assumed, up to 60\% if the average between upper and lower limits is assumed. At this time, the measurement campaign is completed and the analysis is ongoing. The possible resonances at $E_{\rm c.m.}$=68, 100 and 206\,keV, corresponding to the levels at $E_x=8862, 8894$ and 9000\,keV in $^{23}{\rm Na}$ respectively, still remain unobserved.

In order to study the lower energy resonances and the direct capture contribution, a second measurement campaign, using the high-efficiency 4$\pi$ BGO detector, is underway.

\subsection{The $\Mgpg$ reaction}
\label{sub:25Mgpg}
\begin{figure}[!hbt]
\begin{center}
\resizebox{.9\hsize}{!}{\includegraphics{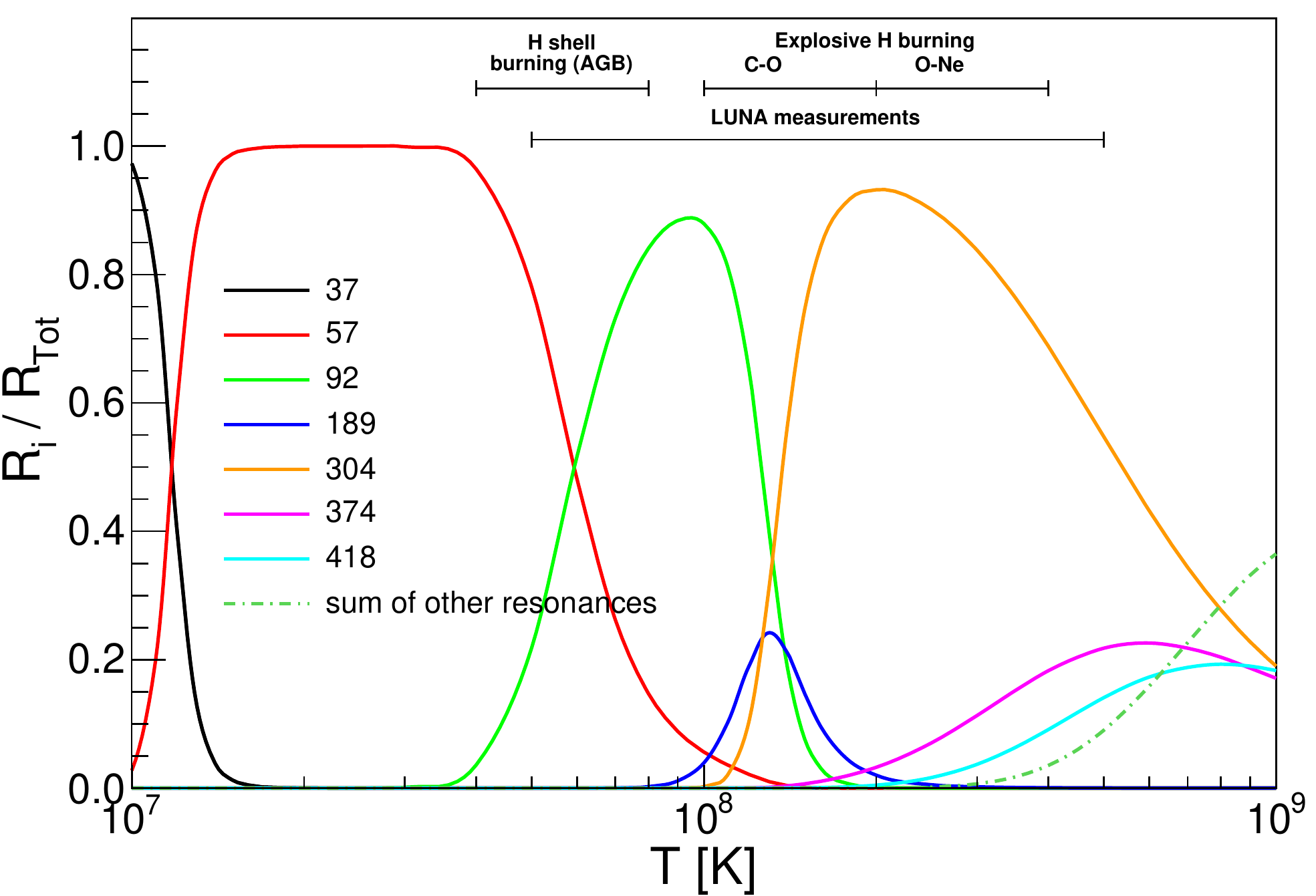}}
\caption{Fractional contribution of resonances to the total reaction rate of $\Mgpg$, as a function of temperature. 
\label{fig:fractionalrate25Mgpg}}
\end{center}
\end{figure}

The rate of the $\Mgpg$ reaction ($Q = 6.306\rm\,MeV$) at astrophysically relevant energies is characterised by several narrow resonances. Some of them have been studied in previous experiments \cite{Champagne1989,Endt1986,Endt1988,Endt1987,Iliadis1990,Powell1998} down to $E_{\rm c.m.} = 189$\,keV. The $\rm^{26}Al$ level structure has states compatible with the existence of additional low-lying resonances at $E_{\rm c.m.} = 37, 57, 92, 108$, and $130\,$keV, that have been identified in indirect experiments through transfer reaction studies \cite[and references therein]{Iliadis1996}. The 92\,keV resonance is thought to be the most important for temperatures ranging from 50 to 120\,MK, as shown in Fig. \ref{fig:fractionalrate25Mgpg}.
The $\Mgpg$ resonances decay through complex $\gamma$-ray cascades either to the 5$^+$ ground state - $^{26}{\rm Al}^g$ - or to the 0$^+$ isomeric state - $^{26}{\rm Al}^m$ - at $E_x = 228\,$keV. The ground state feeding probability $f_0$ is of particular relevance for astronomy. In fact $^{26}$Al in its ground state decays via $\beta^+$ with a half life of 0.7\,My into the first excited state of $^{26}$Mg with a subsequent $\gamma$-ray emission. This $\gamma$-ray is observed by satellite telescopes. On the contrary $^{26}{\rm Al}^m$ $\beta^+$ decays, with a short half life of $\tau_{1/2}^m$ = 6.3\,s, exclusively to the ground state of $^{26}$Mg and therefore does not lead to the emission of $\gamma$-rays. 
Therefore, a precise determination of $f_0$ is essential in order to compare the calculated $^{26}$Al yields with the astronomical observations.

\begin{figure}[!hbt]
\begin{center}
\resizebox{.9\hsize}{!}{\includegraphics{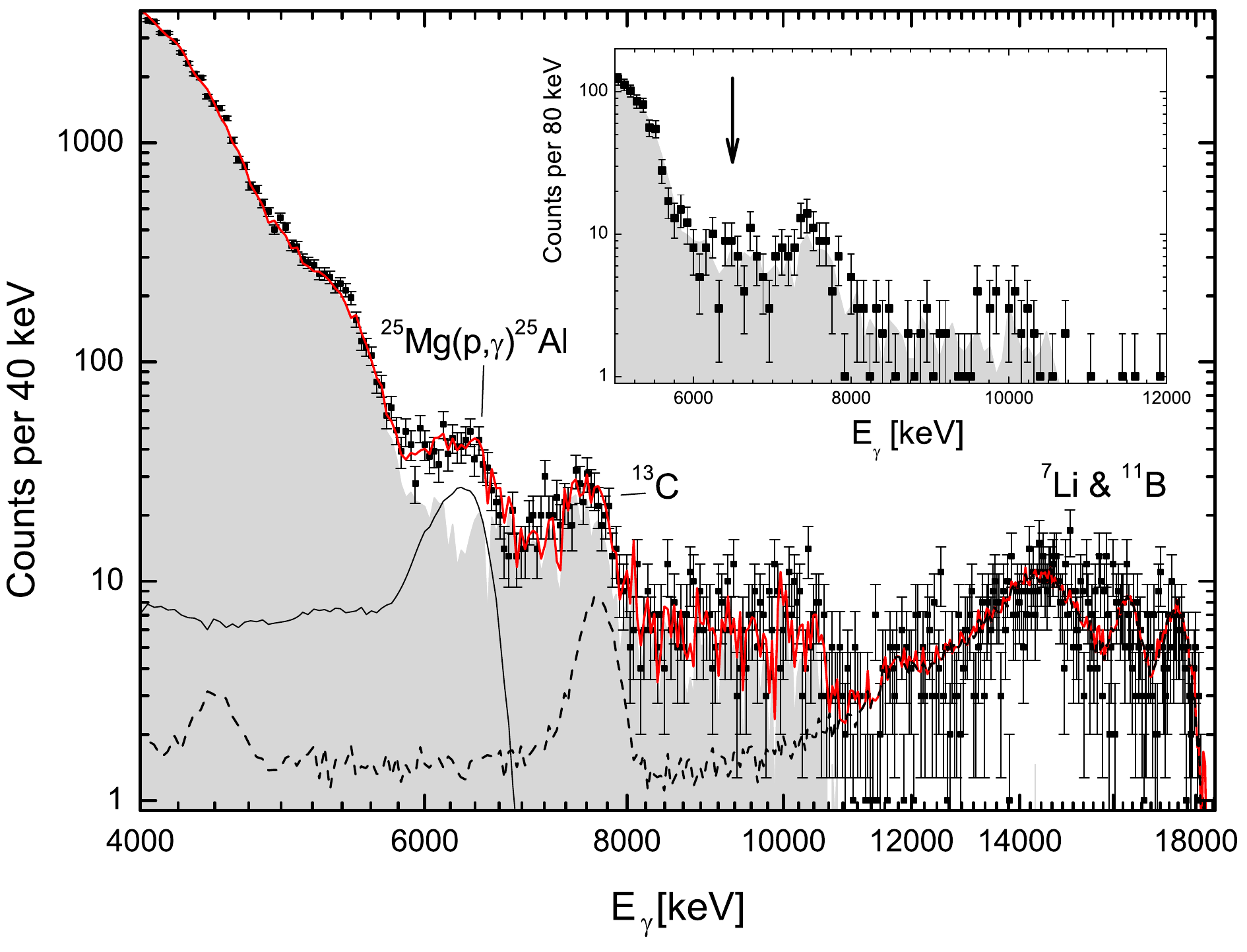}}
\caption{The BGO spectrum at the energy corresponding to the $E_{\rm c.m.}=92\,$keV resonance, as from \cite{Strieder2012}. 
\label{fig:spectrum25Mgpg}}
\end{center}
\end{figure}

Measurements at LUNA were performed using the 4$\pi$ BGO detector. Taking full advantage of the extremely low $\gamma$-ray background, the absolute $\Mgpg$ resonance strengths of the $E_{\rm c.m.}=92, 189$, and 304\,keV resonances have been measured with unprecedented precision and an upper limit was set for the 130\,keV one \cite{Limata2010,Strieder2012}.
In particular, the $92\,$keV resonance was directly observed for the first time \cite{Strieder2012}, see Fig. \ref{fig:spectrum25Mgpg}.
A resonance strength $\omega\gamma=(2.9 \pm 0.6) \cdot 10^{-10}\,$eV was determined. Most of the uncertainty arises from the lack of knowledge of the 92\,keV resonance level scheme.
In spite of tremendous experimental efforts in background reduction and target preparation as well as improvements in $\gamma$-ray detection, other low-energy resonances are still unaccessible to direct detection.

In addition, the 189 and 304\,keV resonances were also studied with a HPGe detector allowing for a precise determination of the corresponding branching ratios \cite{Limata2010}.
Because of the limited $\rm^{25}Mg$ isotopic enrichment of the target, the $E_{\rm c.m.}=214\,$keV resonance in $^{24}{\rm Mg}(p,\gamma)^{25}{\rm Al}$ and the $E_{\rm c.m.}=326\,$keV resonance in $^{26}{\rm Mg}(p,\gamma)^{27}{\rm Al}$ had to be fully characterized \cite{Limata2010}. The strength of the 189\,keV resonance yielded a value in agreement with the BGO measurement \cite{Strieder2012}, but 20\% larger compared to previous works. 

For the $E_{\rm c.m.}=$189 and 304\,keV resonances, $f_0$ could be experimentally determined \cite{Limata2010,Strieder2012}. Due to their weakness, this was not possible for the low energy resonances, therefore for these resonances $f_0$ relies mainly on literature information. The main source is Ref. \cite{Endt1987}, which is to a large extent based on the experimental work published in Ref. \cite{Endt1988}. For resonances at $E_{\rm c.m.}=37$ and 57\,keV the $f_0$ determination seems to be well grounded while for the 92\,keV resonance there is no actual experimental information in \cite{Endt1988}.
Unfortunately, the other literature information about the 92\,keV resonance is contradictory. The measurements performed at LUNA \cite{Strieder2012} suggest a stronger feeding of $^{26}{\rm Al}$ states that predominantly decay to the isomeric state reducing the ground state fraction. Therefore, a $f_0$ of 60$^{+20}_{-10}\,\%$ was recommended \cite{Strieder2012,Straniero2013}.

Furthermore, the $E_{\rm c.m.}=304\,$keV resonance,  was studied with accelerator mass spectrometry (AMS) \cite{Limata2010}. This off-line method provides the cross section to the $^{26}{\rm Al}^g$ by the ultra sensitive measurement of the ratio between the number of $^{26}{\rm Al}$ nuclei produced during the proton irradiation of the $^{25}{\rm Mg}$ targets and a known amount of $^{27}{\rm Al}$ added as a {\it spike} to the samples. Isotopic ratios of the order of $10^{-11}$ could be measured using the CIRCE AMS facility \cite{Terrasi2007} and allowed for a determination of the resonance strength in agreement with the  BGO measurements within 5\%.

On the basis of the LUNA experimental results, the rates of $\Mgpg^g$ and $\Mgpg^m$ were updated. Over the temperature range from 50 to 150\,MK, the rate of the $^{26}{\rm Al}^m$ production was found to be 4 times higher, while the $^{26}{\rm Al}^g$ 20\% higher than previously assumed. At  $T=100\,$MK the revised total reaction rate \cite{Straniero2013} was determined to be a factor of 2 higher than \cite{Iliadis2010_NPA2}.

These new values have several consequences. The expected production of $\rm^{26}Al_{gs}$ in stellar H-burning sites was found to be lower than previously estimated. 
This implies a reduction of the estimated contribution of Wolf-Rayet stars to the galactic production of $\rm^{26}Al$. The deep AGB extra-mixing, often invoked to explain the large excess of $\rm^{26}Al$ in some O-rich grains belonging to pre-solar grains originated in AGB stars, does not appear a suitable solution for $\rm^{26}Al/^{27}Al > 10^{-2}$.

\noindent The substantial increase of the total reaction rate makes the Globular Cluster self-pollution caused by massive AGB stars a more reliable scenario for the reproduction of the Mg-Al anti-correlation \cite{Straniero2013}.

\subsection{Upcoming measurements}
\label{sub:future}
\begin{figure}[!hbt]
\begin{center}
\resizebox{.9\hsize}{!}{\includegraphics{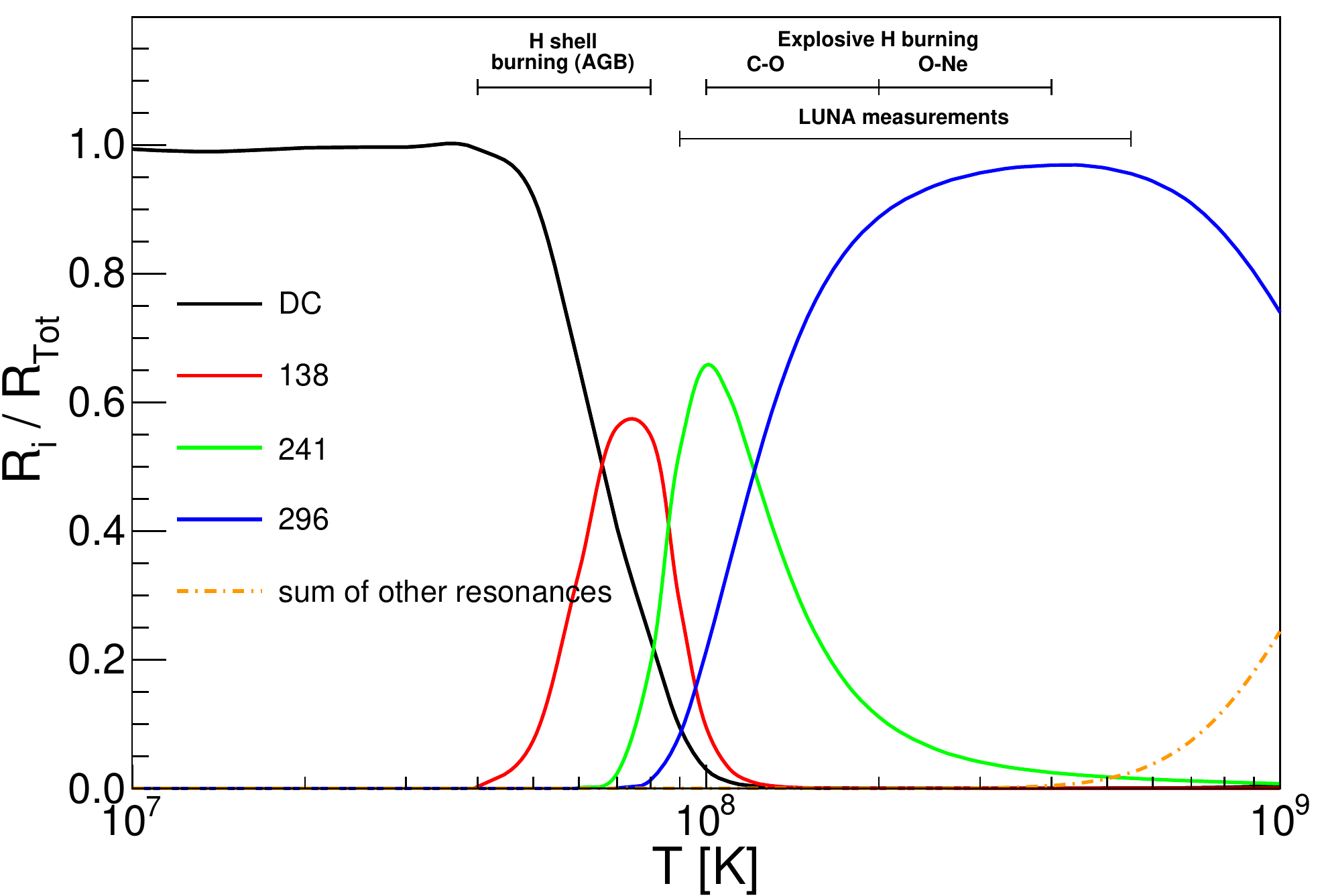}}
\caption{Fractional contributions to the reaction rate of $\Napg$, as a function of temperature, see text for details.
\label{fig:23NaReactionRates}}
\end{center}
\end{figure}

The $\Napg$ reaction links the NeNa and the MgAl cycles, as shown in Fig. \ref{fig:cycles}. Only upper limits could be determined for the strength $\omega\gamma$ of the resonance at $E_\mathrm{c.m.} = 138\,\mathrm{keV}$, which populates the $E_x = 11831\,\mathrm{keV}$ excited state in $^{24}{\rm Mg}$ \cite{Hale2004}. The most recent limit obtained in a direct measurement is 5.17\,neV (95\% confidence level) \cite{Cesaratto2013}. Assuming this value, the resonance dominates the reaction rate over the temperature range $20\lesssim T \lesssim80\,$MK and thus would be most relevant for AGB nucleosynthesis.
In Figure \ref{fig:23NaReactionRates} the fractional reaction rate is presented assuming one tenth of this upper limit value, according to the prescription of \cite{NACRE99-NPA}.
Efforts at LUNA aim at a direct measurement of its strength. Also in this case, the two complementary approaches of high efficiency, with the large BGO detector, and high resolution, with a HPGe detector, will be used. The segmentation of the BGO detector may help to discriminate beam-induced background contributions if the decay schemes of the background reactions are well known. The same principle may aid in the identification of the signal from the studied reaction. Unfortunately the properties and branchings of the level at $E_x=11831\,\mathrm{keV}$ in ${}^{24}$Mg are unknown. Due to the large full energy detection efficiency of the BGO detector, its measurements are less sensitive to uncertainties of the level branchings. On the other hand, gaining information about the branchings from these measurements is very difficult. The opposite is true for the HPGe detector, so that a combination of measurements with both detection techniques may make possible the determination of both the total cross section and branchings.

Another important factor that influences the detection sensitivity of the measurement setup is the background. Environmental background does not reach the energy region of the sum signal at $E_x=11831\,\mathrm{keV}$ for this reaction, except for the muon-induced background which is greatly suppressed at LNGS thanks to its underground location. However cascade transition $\gamma$-rays, to be detected with the HPGe detector, will also have energies below $E_x=11831\,\mathrm{keV}$ and their detection may thus be affected by environmental backgrounds. In this case, a further suppression of environmental $\gamma$-rays can improve the sensitivity of the setup. A standard way to reduce environmental $\gamma$-ray backgrounds is a lead shielding, that in an underground location, is especially effective. In fact, surface experiments benefit of lead shielding only up to a limited thickness, since the secondary radiation produced by the interaction of cosmic muons with the lead limit the maximum useful thickness. On the contrary, in a muon suppressed environment the thickness is limited only by practical reasons, as discussed in another paper of this Topical Issue \cite{Best2016}.
A customized massive lead shielding has been designed and installed recently at the second beam line of LUNA. It consists of few large pieces mounted on rails, allowing a quick and easy access to the target, that needs to be replaced regularly due to beam-induced target degradation. The modular design allows the shielding either to fit the BGO at $0^\circ$, or to fit a HPGe detector facing the target at $55^\circ$ angle with respect to the beam direction. The shielding ensures a minimum thickness of 10\,cm of lead for the BGO and 15\,cm for the HPGe configurations, respectively, reducing significantly the environmental $\gamma$-ray background in the detectors \cite{Boeltzig2016}.

As discussed, this improvement is more beneficial for the measurements planned with the HPGe detector, as $\gamma$-rays from the $\Napg$ reaction may fall in the natural background region. For the BGO setup, it also offers an improvement above the natural $\gamma$-ray background region, as pile-up and random coincidence summing are effectively reduced. 

\section{Summary and outlook}
\label{sec:Summary}
In this review, we presented the contributions made by LUNA experiments in furthering our knowledge in modelling evolution and nucleosynthesis in AGB stars and in Novae explosions.
The experimental program at LUNA with the 400\,kV accelerator will provide data on key reactions of CNO, NeNa and MgAl cycles at least for another 5 years. The upcoming LUNA-MV project will further extend the number of reactions to be studied underground, giving the possibility to cover wide energy ranges in the same Laboratory \cite{Aliotta2016}.

The success of this pioneering project is also proven by the fact that several other underground laboratories, e.g. in USA \cite{DIANA} and China \cite{JUNA}, are planning to install accelerators for the study of nuclear processes relevant to astrophysics.

\section*{Acknowledgments}
Financial support from Ministero dell'Istruzione, dell'Uni\-ver\-si\-t\`a e della Ricerca (MIUR) under the actions FIRB RBFR08549F and PRIN 20128PCN59\_002 is gratefully acknowledged. P.M. acknowledges support from the ERC Consolidator Grant funding scheme ({\em project STARKEY}, G.A. n.615604).

\bibliographystyle{unsrt}
\bibliography{HShellAndExplosiveBurning}

\end{document}